\documentclass[prd,twocolumn,superscriptaddress,floatfix,amsmath,amssymb,amsfonts,longbibliography]{revtex4-2}

\usepackage{float} 
\usepackage{scalerel}
\usepackage[normalem]{ulem}
\usepackage[english]{babel}
\usepackage{graphicx}
\usepackage{dcolumn}
\usepackage{bm}
\usepackage{bbm}
\usepackage{blindtext}
\usepackage{verbatim}
\usepackage{relsize}
\usepackage{mathrsfs}
\usepackage{musicography}
\usepackage{amsmath}
\usepackage{blindtext}
\usepackage{cancel}
\usepackage{physics}
\usepackage{epstopdf}
\usepackage{mathtools}
\usepackage{blindtext}
\usepackage{tensor}
\usepackage[dvipsnames]{xcolor}
\usepackage[usenames,dvipsnames]{pstricks}
\usepackage{epsfig}
\usepackage{pst-grad}
\usepackage{pst-plot} 
\usepackage{hyperref}
\usepackage{verbatim}
\usepackage{slashed}
\usepackage{dsfont}
\usepackage[english]{babel}
\usepackage{amsmath,amssymb,amsthm} 
\usepackage{lipsum}

\newcommand{\mf}{\mathsf}

\newcommand{\ii}{\mathrm{i}}

\newcommand{\tc}[1]{\textsc{#1}}

\newcommand{\gIB}{g_{\scriptscriptstyle IB}}

\newcommand{\xv}{{\bm{x}}}

\newcommand{\lettersec}[1]{\noindent\textbf{\textit{#1---}}}

\allowdisplaybreaks[1] 

\newcommand{\mo}[1]{\textcolor{OrangeRed}{\textit{[M: #1]}}
}

\newcommand{\fra}[1]{\textcolor{Bittersweet}{\textit{[F: #1]}}
}

\newcommand{\srr}[1]{%
\textcolor{red}{\sout{#1}}
}

\begin{document}

\title{
Bose polarons as relativistic Unruh-DeWitt detectors: \\Entanglement harvesting from Bose-Einstein condensates
}

\author{T. Rick Perche}
\email{rick.perche@su.se}

\affiliation{Nordita,
KTH Royal Institute of Technology and Stockholm University,
Hannes Alfvéns väg 12, 23, SE-106 91 Stockholm, Sweden}

\author{Francesco Gozzini}
\email{gozzini@thphys.uni-heidelberg.de}

\affiliation{Kirchhoff-Institut f\"ur Physik, Universit\"at Heidelberg,\\
Im Neuenheimer Feld 227, 69120 Heidelberg, Germany}

\author{Markus K. Oberthaler}
\email{markus.oberthaler@kip.uni-heidelberg.de}

\affiliation{Kirchhoff-Institut f\"ur Physik, Universit\"at Heidelberg,\\
Im Neuenheimer Feld 227, 69120 Heidelberg, Germany}

\begin{abstract}
    We show that a bound impurity in a Bose-Einstein condensate can be directly mapped to an Unruh-DeWitt detector interacting with a relativistic quantum field. We provide explicit experimental parameters for an implementation using ${}^{41}$K impurities coupled to a ${}^{87}$Rb condensate via finite-time Feshbach tuning. 
    As an application, we study the extraction of vacuum entanglement from distant regions of the condensate and find viable parameters for the implementation of entanglement harvesting.
    

    

    
\end{abstract}

\maketitle

\lettersec{Introduction}The advent of quantum field simulators capable of reproducing aspects of quantum field dynamics in curved spacetime has created unprecedented opportunities to test foundational predictions of relativistic quantum field theory (QFT)~\cite{Mo_ViermannNature2022,MO_TajikPNAS2023,MO_SchuetzholdPPNP2025}. 
A growing number of experimental platforms — ranging from ultracold atoms, trapped ions~\cite{Mo_Roos2011RelativisticIons}, photonic~\cite{Mo_Longhi2011OpticalRelativisticReview}, optomechanical ~\cite{MO_optomechanics_metric,MO_acousticquantumvacuum} and superconducting architectures~\cite{Mo_Pedernales2013CircuitQEDRelativistic} to polariton fluids~\cite{MO_Falque2025PolaritonCurvedSpacetimes} — can engineer effective metrics, horizons, and nontrivial expansion dynamics~\cite{Mo_BarceloLRR2011}.

As a result, phenomena long viewed as purely theoretical, such as horizon-induced correlations~\cite{Mo_Shi2023OnChipBlackHole}, particle production in time-dependent backgrounds~\cite{Mo_optics_Steinhauer:2021fhb,Mo_ViermannNature2022}, and entropy measurements in QFT~\cite{AreaLawsBEC2025}, are becoming experimentally accessible. These new experimental capabilities open a path toward using laboratory systems to probe regimes that lie far beyond the reach of astrophysical or cosmological measurements. 

The key feature of all experimental systems so far is the ability to directly detect spatial correlations of the analog relativistic quantum field. However, we are still missing implementations of probes that can couple to the field locally in space and time. Local measurements are essential for probing fundamental features of QFTs that only become evident in finite spacetime regions, such as properties of local algebras~\cite{algebras2023,FewsterAlgebras}, covariant measurement schemes~\cite{FewsterVerch,fewster3}, and more relevant for this work, entanglement between non-complementary regions~\cite{witten,KlcoEntStruc2023,KlcoEntStruc2023II}.

Local probes in QFT are usually implemented as particle detector models, or Unruh-DeWitt (UDW) detectors~\cite{Unruh1976,Unruh-Wald,DeWitt}. These consist of effectively any localized quantum system that couples to a quantum field~\cite{generalPD,QFTPD}. For instance, UDW detectors can be implemented in practice with electro-optic sampling~\cite{Mo_Onoe2022RapidlySwitchedUDW,Mo_Settembrini2022VacuumCorrelationsNatCommun}, superconducting qubits~\cite{adamExperimentalEH2025}, and proposals for interferometric schemes in Bose-Einstein condensates have been put forward~\cite{MO_Gooding2020InterferometricUnruhPRL,Mo_VacuumEntanglement_BEC_Gooding_2024}. 

When coupled for finite times, UDW detectors can implement quantum information protocols that fully utilize the local degrees of freedom of quantum fields~\cite{teleportation,Hotta2011,Jonsson2}. In particular, they can quantify entanglement in the vacuum of a relativistic QFT through entanglement harvesting~\cite{Valentini1991,Reznik2003,Reznik1,Pozas-Kerstjens:2015}. In the protocol, yet to be implemented experimentally, entanglement is extracted from the quantum field by causally disconnected probes.

\begin{figure}[h]
    \centering    \includegraphics[width=0.9\linewidth]{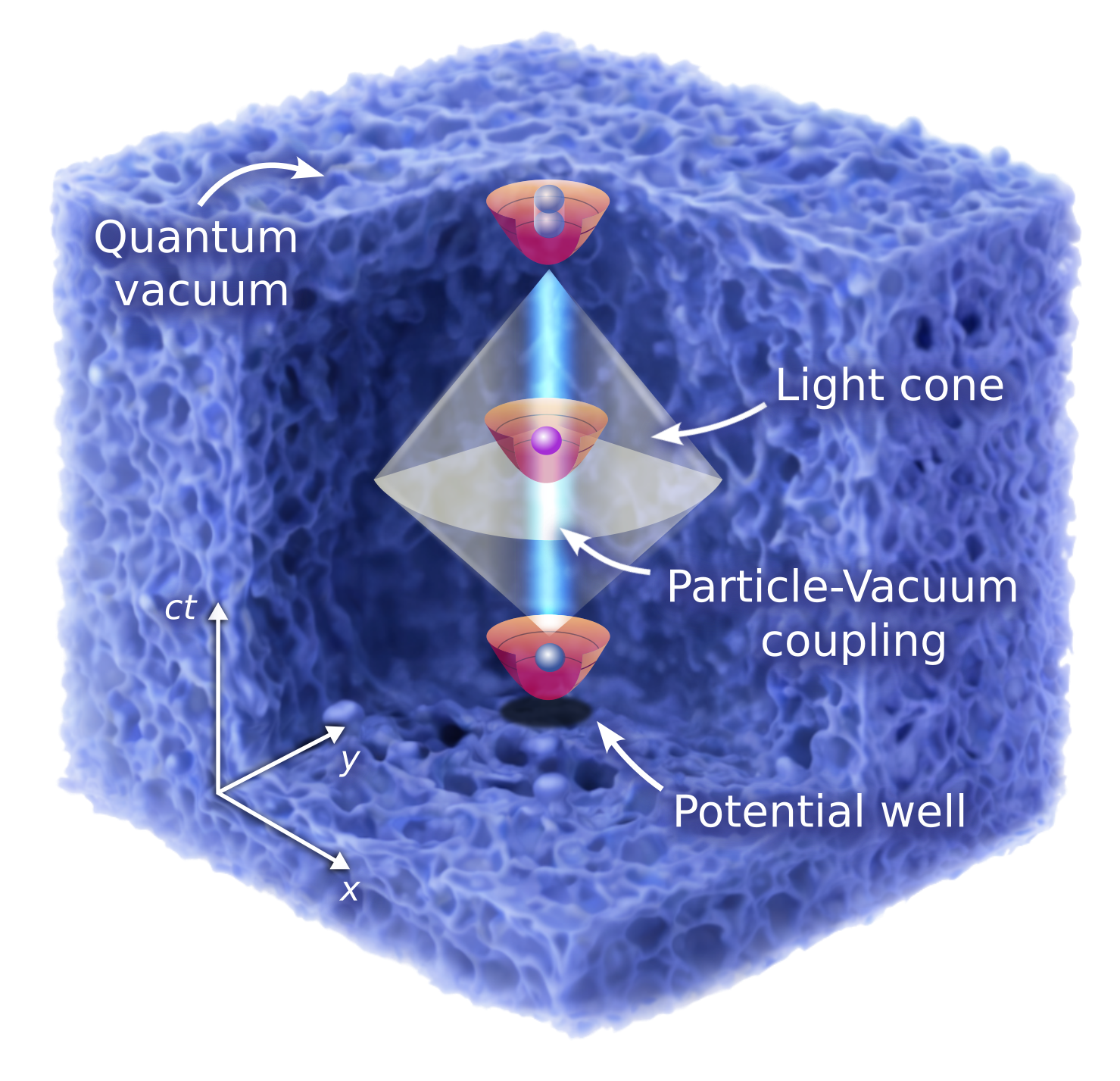}
    \caption{A bound Bose polaron as an Unruh-DeWitt detector. An impurity trapped in a harmonic trap is coupled for a finite time with a Bose-Einstein condensate (BEC) by temporal Feshbach tuning to non-vanishing scattering length. The detector probes local degrees of freedom of the condensate in a causal diamond, determined by the speed of sound. After the interaction, the impurity is entangled with the BEC, capturing  some of its quantum information content. 
    \label{fig:placeholder}    }
\end{figure}

Here, we show that a bound Bose polaron—a trapped impurity immersed in a Bose–Einstein condensate (BEC) \cite{Mo_Grusdt2025RepProgPhys_BosePolaronReview,MO_Massignan2025PolaronsReview,MO_Unruhdetector_with_lattice_Reznik,MO_CasimirIacopo}—realizes a microscopic implementation of an Unruh–DeWitt detector coupled for finite times to the effective relativistic phonon field of the condensate. Ultracold gases offer pristine control over motional degrees of freedom, i.e., spatial mode structure of the detector, and
here we show how one can realize independent temporal control of the impurity–phonon coupling (see also~\cite{PhysRevLett.91.240407,UwePRD2004}, which discusses a related implementation). This realizes a spatiotemporally localized detector suitable for probing quantum-information–theoretic phenomena such as entanglement harvesting. Our microscopic derivation expresses the detector gap, coupling strength, and switching function in terms of experimentally controllable parameters, providing concrete prescriptions for implementing relativistic detector physics in current ultracold atom setups.

\lettersec{Local probes in QFT}The degrees of freedom of a relativistic quantum field theory are spread over spacetime regions that are fully determined by information at a given surface, the so-called causal diamonds~\cite{Haag,advancesAQFT2015,kasiaFewsterIntro}. 
Due to this fact, long-time interactions effectively probe degrees of freedom localized in a spatial region with extension much larger than the probe itself. Local effects in QFT can then only be seen when probing the field for finite times~\cite{quantClass}, for instance using UDW detectors. 


In its simplest form, a UDW detector consists of a two-level system linearly coupled to a free scalar quantum field $\hat{\phi}(\mf x)$ in a background spacetime $\mathcal{M}$. However, here we focus on the case where the detector couples to the conjugate momentum $\hat{\pi}(\mf x) = \partial_t \hat{\phi}(\mf x)$ in 3+1 dimensional Minkowski spacetime~\cite{derivativeJorma,matheusDerivative2023,adamDerivative2024} with inertial coordinates $\mf x = (t,\bm x)$. We also assume that the probe undergoes an inertial trajectory. If the energy gap between the detector's ground and excited states is $\hbar\Omega$, the coupling between the probe and the field $\hat{\pi}(\mf x)$ is described by an interaction Hamiltonian density of the form~\cite{us}
\begin{equation}\label{eq:HIUDW}
    \hat{\mathcal{H}}_I(\mf x) = \lambda ( \Lambda^*(\mf x)e^{\ii \Omega t}\hat{\sigma}^++ \Lambda(\mf x)e^{-\ii \Omega t}\hat{\sigma}^-)\hat{\pi}(\mf x),
\end{equation}
where $\lambda$ is a coupling constant, $\Lambda(\mf x) = \chi(t)f(\bm x)$ is a spacetime smearing function that factors as a (dimensionless) real switching function $\chi(t)$ and a (potentially complex) smearing function $f(\bm x)$ with units of density. $\Lambda(\mf x)$ determines the region of spacetime in which the probe couples to the field, and $e^{\pm \ii \Omega t} \hat{\sigma}^\pm$ are the time-evolved raising and lowering operators in the detector Hilbert space. We use conventions such that $\hat{\pi}^2$ has units of energy density. 

Applications of the UDW model in curved spacetimes and non-inertial trajectories have been proven effective for studying effects such as Hawking radiation~\cite{Hawking1975,JormaHawking,JormaHowThermal} and the Unruh effect~\cite{Unruh1976,Takagi,matsasUnruh,UnruhPhilosophers,mine}. Moreover, the UDW model has been shown to reproduce the essential features of the interactions of quantum fields coupled to any bosonic field, such as an atom or superconducting qubits coupled to electromagnetism~\cite{Pozas2016,Nicho1,richard}, and quantum systems coupled to linearized quantum gravity~\cite{remi,pitelli,boris}. However, the difficulties in implementing a finite-time coupling present a significant challenge to the experimental implementation of UDW detectors~\cite{f1}.


A two-level system that interacts with the field according to Eq.~\eqref{eq:HIUDW} samples field degrees of freedom localized within the causal diamond that contains the region where $\Lambda(\mf x)$ is non-zero, allowing direct access to local observables in QFT~\cite{FewsterVerch}. Indeed, to leading order in the coupling constant $\lambda$, the detector's final state only depends on the two-point function of $\hat{\pi}(\mf x)$ smeared against the region $\Lambda(\mf x)$. Explicitly, the vacuum two-point function of the operator $\hat{\pi}(\mf x)$ can be written as
\begin{equation}\label{eq:pipicorr} 
    \langle\hat{\pi}(\mf x) \hat{\pi}(\mf x')\rangle = \frac{1}{(2\pi)^3}\int \dd^3 \bm k \frac{\hbar \omega_{\bm k}}{2} e^{\ii \mf k \cdot (\mf x - \mf x')},
\end{equation}
where $\mf k \cdot \mf x = - \omega_{\bm k} t + \bm k \cdot \bm x$, $\omega_{\bm k} = c |\bm k|$, and $c$ is the speed of light. The final state of a UDW detector only depends on the two-point function~\eqref{eq:pipicorr} integrated along the support of $\Lambda(\mf x)$.

\lettersec{Phonons as relativistic quantum field} 
Since the seminal paper by Unruh on experimental black hole evaporation~\cite{unruhExperimentalBHE1981}, numerous theoretical and experimental results building on the analogy between propagation of fields in curved spacetimes and propagation of sound waves in fluids have been obtained~\cite{Mo_BarceloLRR2011}.
Especially, atomic superfluids are an ideal platform to implement quantum field simulators, as low-energy phonons behave as a quantum massless scalar field in a Lorentzian metric determined by the background condensate~\cite{MO_Garay2000SonicAnalogPRL, barceloAnalogueGravity2005}. By splitting the atomic quantum field as a classical background and quantum fluctuations, $\hat{\Psi}(\mf x) = \Psi_0(\mf x) + \delta\hat{\Psi}(\mf x)$~\cite{Mo_PitaevskiiStringari2018ReprintBECSuperfluidity}, the BEC density can be written (to leading order in the fluctuations) as
\begin{equation}
    \hat{\rho}_b(\mf x) = \hat{\Psi}^\dagger(\mf x)\hat{\Psi}(\mf x) \approx \rho_0(\mf x) + \delta\hat{\rho}(\mf x).
\end{equation}
Here, $\rho_0 = \Psi_0^\dagger\Psi_0$ is the (classical) background density and $\delta\hat{\rho}(\mf x)$ encodes its quantum fluctuations. 
It can be shown that in the low-energy regime $\delta\hat{\rho}$ behaves analogously to the momentum of a real scalar field (see e.g.~\cite{Mo_TolosaSimeon2022CurvedExpandingBEC} and Appendix~\ref{app:BECQFT}). This quantum field evolves in a spacetime with Lorentzian metric where the effective speed of light is given by the speed of sound in the BEC. The speed of sound in the condensate is given by $c_s^2=\rho_0 g_{bb}/m_b$, where $g_{bb}=4\pi\hbar^2 a_{bb}/m_b$; here $a_{bb}$ is the $s$-wave scattering length characterizing the boson--boson interaction and $m_b$ is the mass of the boson.
In particular, when the density of the condensate is constant $\rho_0(\mf x) = \rho_0$, i.e., $c_\tc{s}(\mf x) = c_\tc{s}$ and the effective spacetime is Minkowski. The density fluctuations $\delta\hat{\rho}(\mf x)$ can then be written as 
\begin{align}
  \delta\hat\rho(\mf x) &= \frac{1}{\sqrt{g_{bb}V}}\sum_{\bm k \neq 0}\sqrt{\frac{\hbar\omega_{\bm k}}{2}}\left( \hat b_{\bm k} e^{\ii \mf k \cdot \mf x} + \hat b_{\bm k}^\dagger e^{- \ii \mf k \cdot \mf x}\right),
\end{align}
with $V$ the normalization volume, $\hat b_{\bm k}, \hat b^\dagger_{\bm k}$ the phonon creation and annihilation operators, \mbox{$\mf k\cdot \mf x = -\omega_{\bm k}t + \bm{k}\cdot \bm{x}$}, and we assume the relativistic dispersion (energy-momentum) relation $\omega_{\bm{k}} \approx c_\tc{s}|\bm{k}|$, which holds in the low energy regime. Notice that the two-point function of the density fluctuations can then be written as
\begin{equation}~\label{eq:rhotwopt}
    \langle \delta\hat{\rho}(\mf x) \delta\hat{\rho}(\mf x')\rangle =  \frac{1}{g_{bb} V} \sum_{\bm k \neq 0} \frac{\hbar\omega_{\bm k}}{2} e^{\ii \mf k \cdot (\mf x - \mf x')},
\end{equation}
analogous to the two-point function of the conjugate momentum operator in Eq.~\eqref{eq:pipicorr} in a finite volume~\cite{f2}, with the additional factor of $g_{bb}^{-1}$. 

\lettersec{Impurities as local probes} 
Immersing a distinguishable particle, typically of a different atomic species, in Bose-Einstein condensates has been studied in detail and is connected to the physics of polarons in solid state physics \cite{MO_Massignan2025PolaronsReview}. Here, we will use the additional control to trap the impurity in a harmonic trap 
giving rise to a bound Bose polaron. This has already been studied experimentally with a mixture of lithium (impurity) and sodium (BEC), revealing decoherence of motional degrees of freedom~\cite{MO_Scelle2013MotionalCoherence} and the analog of the Lamb shift due to the coupling to the phononic quantum field~\cite{MO_phononic_lamb_shift}. In the weakly interacting limit, which is the regime we are exploiting here, the Bose polaron is know as the Fröhlich polaron~\cite{frolich1954, cucchiettiStrongCouplingPolaronsDilute2006, tempereFeynmanPathintegralTreatment2009}.  
Effectively, the impurity interaction with the BEC is a density-density coupling, controlled by the parameter $g_{ab} = 2\pi\hbar^2 a_{ab}/\mu_{ab}$. Here $\mu_{ab} = m_{a}m_{b}/(m_{a} + m_{b})$ is the reduced mass, $m_a$ is the mass of the impurity atom, and $a_{ab}$ is the interspecies s-wave scattering length.  
%
%
Explicitly, we can write the interaction Hamiltonian density (in the interaction picture) as
\begin{equation}\label{eq:Hintpolaron}
    \hat{\mathcal{H}}_I(\mf x) =  g_{ab} \hat \rho_a(\mf x) \hat \rho_b(\mf x),
\end{equation}
where $\hat \rho_a(\mf x)$ is the time-evolved density of the impurity.
 Denoting the localized eigenstate of the impurity atom with energy $E_{\bm n} = \hbar\Omega_{\bm n}$ by $\ket{\bm n}$ and their wavefunctions by $\psi_{\bm n}(\bm x) = \braket{\bm x}{\bm n}$, we can write   
%
\begin{equation}\label{eq:rhoI-sum}
    \hat{\rho}_a(\mf x) = \sum_{\bm{nm}} \psi_{\bm n}^*(\bm x)\psi_{\bm m} (\bm x) e^{i (\Omega_n - \Omega_m)t}\ket{\bm n}\!\!\bra{\bm m}.
\end{equation}
From now on we assume that the only relevant energy levels in Eq.~\eqref{eq:rhoI-sum} are its ground ($g$) and excited ($e$) states. 
The two-level impurity field operator then becomes
\begin{equation}\label{eq:rhoI-2level}
    \begin{split}
        \hat\rho_a({\mf{x}}) &= e^{i\Omega t} (\psi_e^*\psi_g)(\xv)\hat\sigma^+ + e^{-i\Omega t} (\psi_g^*\psi_e)(\xv)\hat\sigma^- \\
        &+ |\psi_e(\xv)|^2 \hat\sigma^+\hat\sigma^- + |\psi_g(\xv)|^2 \hat\sigma^-\hat\sigma^+
    \end{split}
\end{equation}
where $\Omega = \Omega_e - \Omega_g$ and $\hat{\sigma}^\pm$ are the raising and lowering operators between the states $\ket{g}$ and $\ket{e}$. Since the last two diagonal terms in~\eqref{eq:rhoI-2level} are projectors and do not contribute to the excitation probability of the impurity~\cite{Unruh-Wald}, we drop them in the following. Analogously, in~\eqref{eq:Hintpolaron} we can replace $\hat{\rho}_b(\mf x)$ with $\delta\hat{\rho}(\mf x)$, as $\rho_0$ is a classical term proportional to the identity. Remarkably, the polaron model also allows for a time-varying coupling $g_{ab}(t)$ to be implemented, allowing one to write the interaction Hamiltonian as
\begin{equation}\label{eq:intHamiltonian}
    \hat{\mathcal{H}}_I(\mf x) = \bar{g}_{ab} \chi(t) \left( f^*(\xv)e^{i\Omega t}\hat\sigma^+ +  f(\xv)e^{-i\Omega t}\hat\sigma^- \right) \delta \hat{\rho}(\mf x) \,,
\end{equation}
where  $f(\xv) = (\psi_g^*\psi_e)(\xv)$ and we write $g_{ab}(t) = \bar{g}_{ab}\chi(t)$, factoring out the time dependence in a dimensionless switching function $-1\leq \chi(t) \leq 1$. Comparing~\eqref{eq:intHamiltonian} with~\eqref{eq:HIUDW} shows that the polaron is effectively a UDW detector for the phonon density field. Introducing a timescale $T$ that controls the switching of the interaction through $\chi(t) = \beta(t/T)$, the momentum-coupled UDW detector can be mapped to the polaron model by
\begin{align}
    \hat{\pi}(\mf x)&\mapsto \sqrt{\frac{c_\tc{s}}{c}}\sqrt{g_{bb}}\delta\hat{\rho}(\mf x),\quad  
     \lambda \mapsto \sqrt{\frac{c}{c_\tc{s}}}\frac{\bar{g}_{ab}}{\sqrt{g_{bb}}},
\end{align}
and the relevant rescalings $\Omega \mapsto {\frac{c}{c_\tc{s}}}\Omega$ and  ${T\mapsto \frac{c_\tc{s}}{c} T}.$

Arguably, the most important feature of an UDW detector is to allow finite-time coupling with the field. The control of $g_{ab}(t)$ is achieved by magnetically tuned Feshbach resonances~\cite{MO_Feshbach_RevModPhys.82.1225}. Most distinct pairs of atomic species allow a value of the magnetic field $B_0$ such that scattering length $a_{ab}(B_0) = 0$, implying $g_{ab} = 0$, thus decoupling the impurity and the BEC. Applying a magnetic pulse to the system by varying the external magnetic field $B(t)$ away from $B_0$, we can implement a finite time coupling, in perfect analogy with the detector model in Eq.~\eqref{eq:HIUDW}.

We can explicitly see how the impurity can measure localized field observables through a finite-time interaction by noticing that its vacuum excitation probability (see Appendix~\ref{app:excprob}) is given by
\begin{equation}
    {\mathcal{P}} = \frac{\bar{g}_{ab}^2}{\hbar^2} \langle \delta \hat{\rho}(\Lambda^-)\delta\hat{\rho}(\Lambda^+)\rangle,
\end{equation}
where $\delta\hat{\rho}(\Lambda^\pm) \equiv \int \dd^4\mf x\, \Lambda^\pm(\mf x) \delta \hat{\rho}(\mf x)$ are the smeared field observable localized by the spacetime functions $\Lambda^-(\mf x) = \chi(t)f(\bm x) e^{-i \Omega t}$ and $\Lambda^+(\mf x) = \Lambda^-(\mf x)^*$. The shape of the wavefunction $(\psi_g^*\psi_e)(\bm x)$ determines the spatial localization of the observables probed, and $\chi(t)$ their extension in time, encoding expected values of local operators of algebraic QFT in directly accessible physical quantities.


\lettersec{Experimental platform}The experimental system of ultracold mixtures offers a broad range of control parameters, which we will discuss in the context of UDW detectors and relativistic quantum fields. As a very suitable combination of species, we propose a potassium–rubidium mixture \cite{FirstKRb2016}, where species-selective trapping of ${}^{41}$K in a light-shift potential realizes the trapped impurity, while ${}^{87}$Rb acts as the substrate for the phononic relativistic quantum field \cite{barceloAnalogueGravity2005}. The large fine-structure splitting of rubidium allows for the realization of species selective dipole potentials utilizing the existence of a tune-out wavelength \cite{MO_tuneoutwavelength}, i.e., the light-shift potential is only seen by potassium atoms. The trap frequencies, realized with optical tweezers \cite{MO_Endres2016OneDimensionalTweezerArray,MO_Kaufman2021QuantumScienceTweezers} or large-period optical lattices \cite{MO_lattices_RevModPhys}, can be tuned from a few hundred hertz to several tens of kilohertz.

Additionally, the $s$-wave scattering length for rubidium is $a_{bb}=100.4(1)\,a_0$, where $a_0$ is the Bohr radius, combined with the low three-body loss rate, allows for the preparation of large and high-density Bose--Einstein condensates 
\cite{Mo_Streed2006LargeAtomNumberBEC}. For the reported peak density $\rho_0 = 5\times10^{14}\,\mathrm{atoms}\,\mathrm{cm}^{-3}$, the speed of sound is $c_\tc{s}\sim 4.2\,\mathrm{mm/s}$ and the healing length is $\xi\sim 120\,\mathrm{nm}$. Length scales larger than $\xi$ are well captured by the relativistic (phononic) approximation, a prerequisite for the analogy to relativistic quantum field theory. Additionally, temperatures of the order of nK ensure that the zero temperature approximation is valid for short temporal coupling.


The interaction between rubidium and potassium can be controlled via Feshbach resonances employing controlled magnetic fields \cite{MO_KRbfeshbach_first,MO_KRbFeshbach_2008}. For our purposes, the zero crossing of the interaction, i.e., decoupling the detector from the relativistic scalar field, is important and has been studied in detail in the past \cite{MO_zero_crossing_KRb,MO_KRbfeshbach_precise}. Additionally, the interaction can be changed by small variations of the magnetic field ($\sim 1\,\mathrm{G}$) to values as high as $a_{ab} = \pm 1000\,a_0$ \cite{MO_RbK_loss,MO_RbK_loss_remeasured}. 
The scattering properties of rubidium are not changed when varying the magnetic field in this range, ensuring that the substrate for the phononic relativistic quantum field remains unchanged in the process of coupling the detector. 

\lettersec{Entanglement harvesting from a BEC} When systems interact with a quantum field, there are two ways in which they can become entangled. The first (and most commonly considered) method is when the probes exchange information, and entanglement is mediated by the field. The second method for a field to entangle two probes is when the probes extract entanglement from the field itself, effectively performing an entanglement swap operation. These methods contribute differently in different regimes for the interaction between probes and field: in the long time limit, entanglement extraction is negligible compared to entanglement mediated through the field. On the other hand, in the short time limit, when the probes effectively cannot communicate, the only way they can become entangled is by extracting previously existing entanglement in the field (see Fig.~\ref{fig:situation}). This can be used to infer the existence of entanglement between the finite regions that the probes couple to, corresponding to the protocol of entanglement harvesting~\cite{Valentini1991,Reznik2003,Pozas-Kerstjens:2015}.

When two bound impurities (labelled A and B) initialized in their ground state interact with a BEC for finite times, their final state can be written as 
\begin{equation}
    \hat{\rho}_{\tc{ab}} = \begin{pmatrix}
        1 - {\mathcal{P}}_\tc{aa} - {\mathcal{P}}_\tc{bb} & 0 & 0 & \mathcal{M}^*\\
        0 & {\mathcal{P}}_{\tc{bb}} & {\mathcal{P}}_{\tc{ab}} & 0\\
        0 & {\mathcal{P}}_{\tc{ab}}^* & {\mathcal{P}}_\tc{aa} & 0\\
        \mathcal{M} & 0 & 0 & 0
    \end{pmatrix} + \mathcal{O}(g_{ab}^4),
\end{equation}
where
\begin{align}
    {\mathcal{P}}_{\tc{ij}}  &= \frac{\bar{g}_{ab}^2}{\hbar^2}\!\! \int\! \dd^4 \mf x \dd^4 \mf x' \Lambda_\tc{i}(\mf x) \Lambda_\tc{j}(\mf x') e^{- \ii \Omega(t-t')}\langle\delta\hat{\rho}(\mf x)\delta\hat{\rho}(\mf x')\rangle,\nonumber\\
    \mathcal{M} &= -\frac{\bar{g}_{ab}^2}{\hbar^2} \!\!\int\! \dd^4 \mf x \dd^4 \mf x' \Lambda_\tc{a}(\mf x) \Lambda_\tc{b}(\mf x') e^{\ii \Omega(t+t')}\langle\mathcal{T}(\delta\hat{\rho}(\mf x)\delta\hat{\rho}(\mf x'))\rangle,
\end{align}
$\tc{I}, \tc{J} \in \{\tc{A},\tc{B}\}$, $\Lambda_\tc{i}(\mf x) = \chi(t)f_\tc{i}(\bm x)$ are the spacetime smearing functions of the interactions, where $\chi(t)$ is determined by the time profile of the magnetic field applied to the system (Feshbach tuning) and $f_\tc{i}(\bm x)$ are determined by the trapping potentials. The entanglement in the final state of the probes can be quantified through the negativity~\cite{VidalNegativity}, a faithful entanglement quantifier for systems of two qubits. Assuming identical shapes for the detectors, ${\mathcal{P}}_\tc{aa} = {\mathcal{P}}_\tc{bb} = {\mathcal{P}}$, the negativity can be written as
\begin{equation}\label{eq:neg}
    \mathcal{N} = \max(|\mathcal{M}| - {\mathcal{P}},0).
\end{equation} 
Overall, the negativity is a competition between a non-local term $\mathcal{M}$ and local noise term ${\mathcal{P}}$, corresponding to the excitation probability of the detectors.

{We now connect theoretical concepts from entanglement quantification in quantum fields to relativistic quantum computing~\cite{Martin-Martinez_Aasen_Kempf_2013,Bruschi_Dragan_Lee_Fuentes_Louko_2013,Martín-Martínez_Sutherland_2014,Layden_Martin-Martinez_Kempf_2016,phil2025} directly to experimentally accessible parameters of our platform. Estimates use analytical predictions within Gaussian approximations, with switching function $\chi(t)=e^{-t^2/2T^2}$ and spatial smearing $f(\bm x)$ approximated by a normal distribution of width $\sigma=\sqrt{\hbar/m_a\Omega}$, where $\Omega$ is the impurity trap frequency (see Appendix~\ref{app:computations}).

\begin{figure}[h!]
    \centering
    \includegraphics[width=0.95\linewidth]{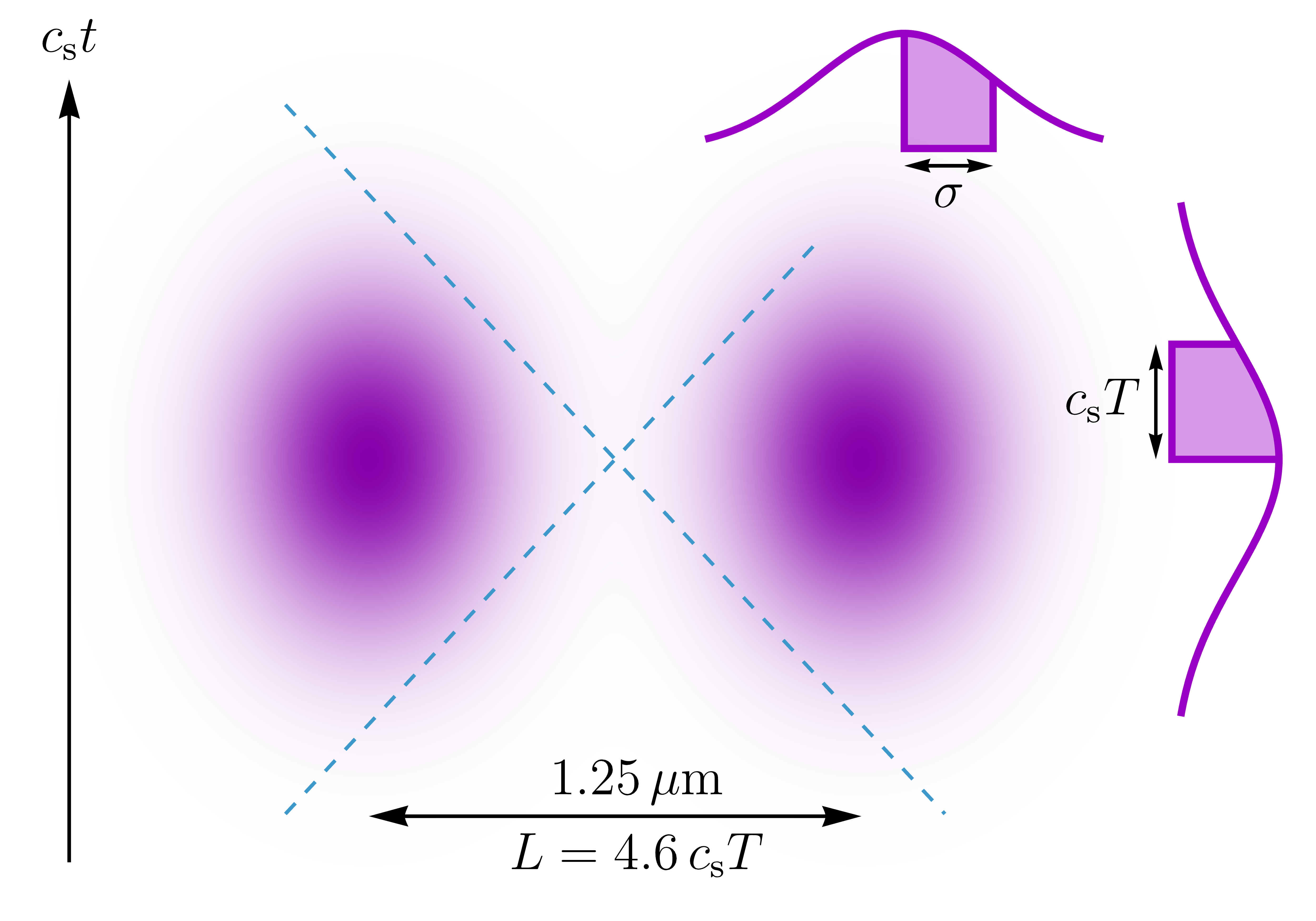}
    \caption{Spacetime density plot of the interaction regions $\Lambda_\tc{i}(\mf x) = \chi(t) f_\tc{i}(\bm x)$ for the probes. In the plot above we use $\sigma = 0.83 c_\tc{s} T$, corresponding to $\Omega = 26.3 \,\text{krad/s}$ when using ${}^{41}$K.   
    }
    \label{fig:situation}
\end{figure}

Since the negativity scales as $a_{ab}^2$, maximizing the interspecies scattering length enhances the signal. The signal is also favored by a large sound speed in the rubidium background gas, requiring high density $\rho_0$. However, three-body losses scale as $a_{ab}^4\rho_0^2$ and deplete the impurities. We therefore fix $\rho_0=5\times10^{14}\,\mathrm{atoms\,cm^{-3}}$, yielding $c_s=4.2\,\mathrm{mm/s}$.
For coupling strength between detector and phononic field we assume $a_{ab}=-1000 a_0$. For the optimal coupling times we find that the loss of the impurity due to three particle collisions is less than $20\%$.

\begin{figure}[h!]
    \centering
    \includegraphics[width=0.99\linewidth]{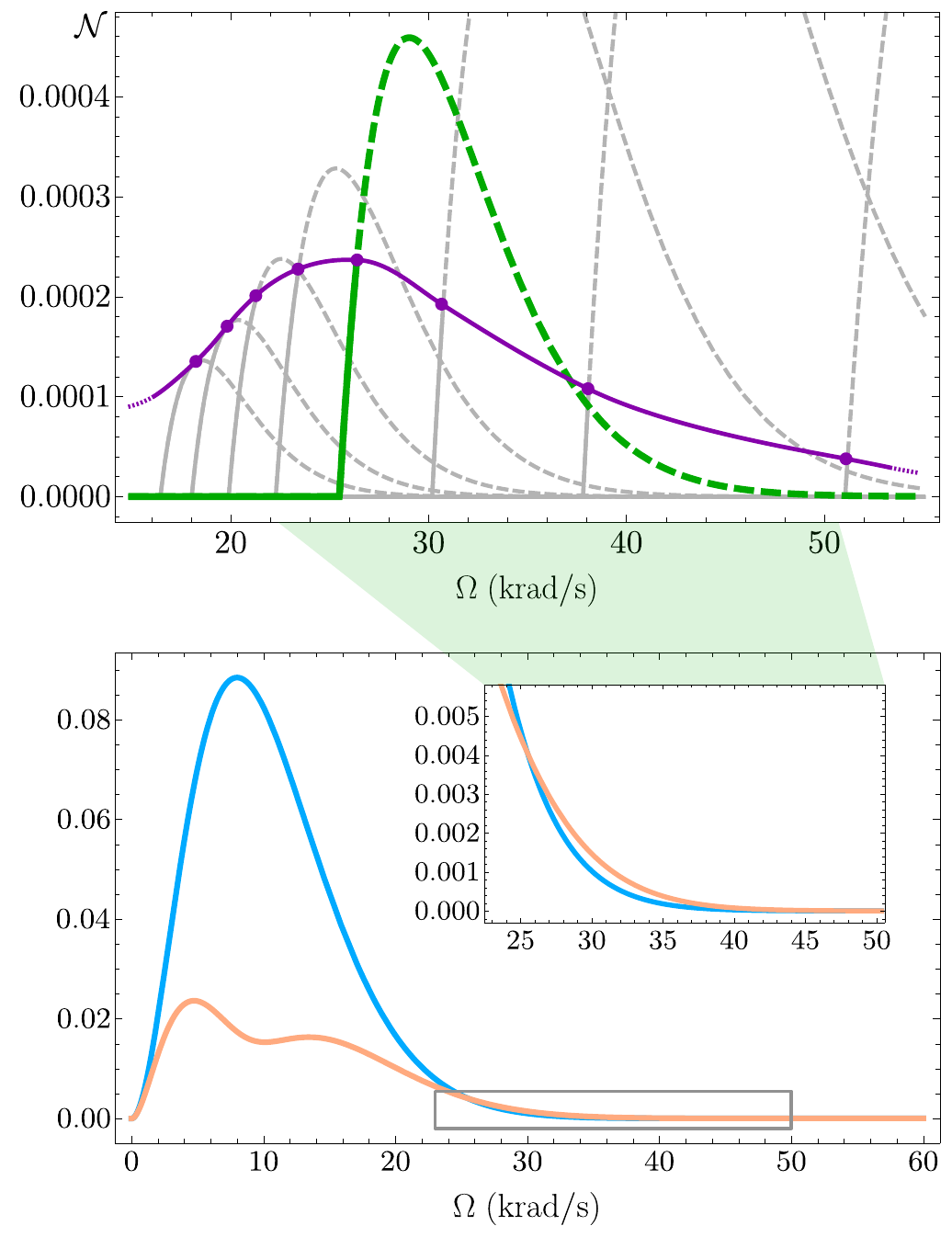}
    \caption{Entanglement harvesting from a Rb BEC with speed of sound $c_\tc{s}= 4.2\,\mathrm{mm/s}$ using ${}^{41}\text{K}$ probes with an interspecies scattering length of $-1000\,a_0$. The upper panel shows the negativity $\mathcal{N} = \max(|\mathcal{M}| - \mathcal{P}, 0)$ as function of the impurity-trap frequency $\Omega$ for a distance of $L_{sep}=4.6\times c_s T$ and interaction time $T$ uniformly decreasing from $T = 105\mu\text{s}$ (leftmost gray curve) to $T = 35\mu\text{s}$ (rightmost gray curve) in steps of $10\mu\text{s}$. The negativity plot with $T = 65\mu\text{s}$ is highlighted in green. The solid parts of the lines have signalling estimator equal to zero. The purple curve interpolates the corresponding maximal negativity values, showing the maximal entanglement that can be harvested. In the lower panel the function $\mathcal{P}$ (blue) and $|\mathcal{M}|$ (orange) for the case $T = 65 \,\mu\mathrm{s}$ are depicted.
    }
    \label{fig:plot_merged}
\end{figure}

In the upper panel of Fig.~\ref{fig:plot_merged} we show the negativity~\eqref{eq:neg} for the chosen atomic species versus trap frequency $\Omega$ and interaction time $T$, with detector separation fixed at $L_{sep}=4.6 \times c_sT \sim 1.25~\mu\text{m}$. The parameters corresponding to solid lines indicate that the signaling estimator~\cite{ericksonNew} vanishes (see Appendix~\ref{app:checks}), confirming that the entanglement is harvested from the BEC. 

The maximal negativity with negligible signalling $\mathcal{N}\sim2\times10^{-4}$ at $\Omega = 26.3~\text{krad/s}$ requires estimating $\mathcal{P}$ and $\mathcal{M}$ with comparable precision; both are about one order of magnitude larger than $\mathcal{N}$, see the lower panel in Fig.~\ref{fig:plot_merged}. While $\mathcal{P}$ follows directly from local detection of excited-state populations, estimating $\mathcal{M}$ requires additional unitary transformations, achievable by controlling the trap potentials before population detection (see Appendix~\ref{app:Mterm})~\cite{MO_phononic_lamb_shift}. For the potentials detailed in Appendix~\ref{app:experimental}, the corresponding infidelities, i.e., systematic deviations of the estimated $\mathcal{P}$ and $\mathcal{M}$ from the ground truth, can be realized below $10^{-5}$, more than one order of magnitude below the expected signal. Thus, the ultimate precision is statistics-limited: for $\mathcal{P}\approx10^{-3}$, see inset Fig.~\ref{fig:plot_merged}, resolving the effect requires more than $10^5$ experimental realizations, straightforwardly obtained in modern experiments~\cite{MO_Strobel2014FisherNonGaussianSpin}.

Furthermore, our platform also allows the implementation of recently proposed routes to enhance harvested entanglement, i.e., increase negativity, including optimized switching functions~\cite{marcos} and superadditivity of bipartite entanglement with multiple probes~\cite{sergi}, each shown to increase the negativity by a factor $10^2$. In particular, the optimization prescribed in~\cite{marcos} applies directly, since arbitrary $\chi(t)$ can be implemented by controlling the magnetic-pulse shape.
}

\lettersec{Conclusions}We show that a bound impurity coupled to a BEC via controlled interspecies interactions directly realizes a Unruh-DeWitt detector interacting with a relativistic field. We found this brings entanglement harvesting, a paradigmatic relativistic quantum-information protocol, within experimental reach even without specific optimization. Due to the pristine control over the spatial and temporal degrees of freedom, this BEC particle detector can be employed for probing local physics in relativistic scenarios not yet experimentally explored. This paves the way for implementations of protocols such as quantum collect calling~\cite{Jonsson2} and quantum energy teleportation~\cite{teleportation,Hotta2011,borisQET2025}, that have their foundation on the underlying causal structure of spacetime. Recently, in~\cite{tian2025} it was also shown that the platform of ultracold dipolar gases allows the study of entanglement harvesting in Lorentz-violating situations.

Our proposal has three main advantages when compared to previous entanglement harvesting experimental proposals in BECs~\cite{Mo_VacuumEntanglement_BEC_Gooding_2024}. First, our proposal makes spacetime locality explicit, allowing controllable qubits to extract and store local information in QFT. Second, it allows for the implementation of oscillatory switching functions, which have been shown to enhance entanglement harvesting by several orders of magnitude~\cite{Reznik1,marcos}. And third, it can be implemented in an effective 3+1 dimensional spacetime. Beyond controlling the probes, we can also envision  pushing the relativistic field to theoretically challenging regimes, for instance by implementing self-interactions. The outcomes of these experiments can give an alternative input for tackling these problems, opening the doors to phenomenological approaches to relativistic quantum information.






\acknowledgements
  
The authors thank Marcos Morote Balboa for verifying computations. We thank Matteo Zaccanti for a crucial information about Feshbach resonances and also Helmut Strobel, Brian Bostwick, Fabian Isler, Alberto Sartori for scrutinizing the experimental feasibility, and Iacopo Carusotto, Silke Weinfurtner, Patricia Ribes-Metidieri, and Sergi Nadal-Gisbert for valuable comments that improved the manuscript. TRP is thankful for financial support from the Olle Engkvist Foundation (no.225-0062). Nordita is partially supported by Nordforsk. This work was made possible by the Deutsche Forschungsgemeinschaft (DFG, German Research Foundation) under Germany’s Excellence Strategy EXC 2181/1 - 390900948 (the Heidelberg STRUCTURES Excellence Cluster).

\bibliography{references}

@misc{f1,
    title = {Notice that the finite coupling required by a detector of this type goes beyond probing finite pulses---it requires a finite time coupling even with the vacuum of the quantum field.}
}

@misc{f2,
    title = {In the infinite volume limit, one has $1/{V} \sum_{\bm k} \mapsto 1/(2\pi)^3\int \dd^3 \bm k$.}
}

@article{Mo_ViermannNature2022,
  author  = {Viermann, Celia and Sparn, Marius and Liebster, Nikolas and Hans, Maurus and Kath, Elinor and Parra-Lopez, Alvaro and Tolosa-Simeon, Mireia and Sanchez-Kuntz, Natalia and Haas, Tobias and Strobel, Helmut and Floerchinger, Stefan and Oberthaler, Markus K.},
  title   = {Quantum field simulator for dynamics in curved spacetime},
  journal = {Nature},
  year    = {2022},
  volume  = {611},
  number  = {7935},
  pages   = {260--264},
  doi     = {10.1038/s41586-022-05313-9}
}

@article{MO_TajikPNAS2023,
  author  = {Tajik, Mohammadamin and Gluza, Marek and Sebe, Nicolas and Sch{\"u}ttelkopf, Philipp and Cataldini, Federica and Sabino, Jo{\~a}o and M{\o}ller, Frederik and Ji, Si-Cong and Erne, Sebastian and Guarnieri, Giacomo and Sotiriadis, Spyros and Eisert, Jens and Schmiedmayer, J{\"o}rg},
  title   = {Experimental observation of curved light-cones in a quantum field simulator},
  journal = {Proceedings of the National Academy of Sciences},
  year    = {2023},
  volume  = {120},
  number  = {21},
  pages   = {e2301287120},
  doi     = {10.1073/pnas.2301287120}
}

@article{MO_SchuetzholdPPNP2025,
  author  = {Sch{\"u}tzhold, Ralf},
  title   = {Ultra-cold atoms as quantum simulators for relativistic phenomena},
  journal = {Progress in Particle and Nuclear Physics},
  year    = {2025},
  volume  = {145},
  pages   = {104198},
  doi     = {10.1016/j.ppnp.2025.104198}
}

@article{Mo_BarceloLRR2011,
  author  = {Barcel{\'o}, Carlos and Liberati, Stefano and Visser, Matt},
  title   = {Analogue Gravity},
  journal = {Living Reviews in Relativity},
  year    = {2011},
  volume  = {14},
  number  = {1},
  pages   = {3},
  doi     = {10.12942/lrr-2011-3}
}

@article{Mo_Roos2011RelativisticIons,
  title   = {Quantum simulation of relativistic quantum physics with trapped ions},
  author  = {Roos, C. F. and Gerritsma, R. and Kirchmair, G. and Z{\"a}hringer, F. and Solano, E. and Blatt, R.},
  journal = {Journal of Physics: Conference Series},
  volume  = {264},
  pages   = {012020},
  year    = {2011},
  doi     = {10.1088/1742-6596/264/1/012020}
}

@article{Mo_Longhi2011OpticalRelativisticReview,
  title   = {Classical simulation of relativistic quantum mechanics in periodic optical structures},
  author  = {Longhi, Stefano},
  journal = {Applied Physics B},
  volume  = {104},
  number  = {3},
  pages   = {453--468},
  year    = {2011},
  doi     = {10.1007/s00340-011-4628-7}
}

@article{Mo_Pedernales2013CircuitQEDRelativistic,
  title   = {Quantum simulations of relativistic quantum physics in circuit {QED}},
  author  = {Pedernales, J. S. and Di Candia, R. and Ballester, D. and Solano, E.},
  journal = {New Journal of Physics},
  volume  = {15},
  pages   = {055008},
  year    = {2013},
  doi     = {10.1088/1367-2630/15/5/055008}
}

@article{Mo_Shi2023OnChipBlackHole,
  title   = {Quantum simulation of {Hawking} radiation and curved spacetime with a superconducting on-chip black hole},
  author  = {Shi, Y. H. and Yang, R. Q. and Xiang, Z. and others},
  journal = {Nature Communications},
  volume  = {14},
  pages   = {3263},
  year    = {2023},
  doi     = {10.1038/s41467-023-39064-6}
}

@article{Mo_optics_Steinhauer:2021fhb,
    author = "Steinhauer, Jeff and Abuzarli, Murad and Aladjidi, Tangui and Bienaim{\'e}, Tom and Piekarski, Clara and Liu, Wei and Giacobino, Elisabeth and Bramati, Alberto and Glorieux, Quentin",
    title = "{Analogue cosmological particle creation in an ultracold quantum fluid of light}",
    eprint = "2102.08279",
    archivePrefix = "arXiv",
    primaryClass = "cond-mat.quant-gas",
    doi = "10.1038/s41467-022-30603-1",
    journal = "Nature Commun.",
    volume = "13",
    pages = "2890",
    year = "2022"
}

@article{MO_Gooding2020InterferometricUnruhPRL,
  title   = {Interferometric {Unruh} Detectors for {Bose-Einstein} Condensates},
  author  = {Gooding, Cisco and Biermann, Steffen and Erne, Sebastian and Louko, Jorma and Unruh, William G. and Schmiedmayer, J{\"o}rg and Weinfurtner, Silke},
  journal = {Physical Review Letters},
  volume  = {125},
  number  = {21},
  pages   = {213603},
  year    = {2020},
  doi     = {10.1103/PhysRevLett.125.213603},
  eprint  = {2007.07160},
  archivePrefix = {arXiv},
  primaryClass  = {gr-qc}
}

@article{Mo_VacuumEntanglement_BEC_Gooding_2024,
doi = {10.1088/1367-2630/ad8675},
url = {https://doi.org/10.1088/1367-2630/ad8675},
year = {2024},
month = {oct},
publisher = {IOP Publishing},
volume = {26},
number = {10},
pages = {105001},
author = {Gooding, Cisco and Sachs, Allison and Mann, Robert B and Weinfurtner, Silke},
title = {Vacuum entanglement probes for ultra-cold atom systems},
journal = {New Journal of Physics},
}

@article{Mo_Onoe2022RapidlySwitchedUDW,
  title   = {Realizing a rapidly switched {Unruh-DeWitt} detector through electro-optic sampling of the electromagnetic vacuum},
  author  = {Onoe, Sho and Guedes, Thiago L. M. and Moskalenko, Andrey S. and Leitenstorfer, Alfred and Burkard, Guido and Ralph, Timothy C.},
  journal = {Physical Review D},
  volume  = {105},
  number  = {5},
  pages   = {056023},
  year    = {2022},
  doi     = {10.1103/PhysRevD.105.056023},
  eprint  = {2103.14360},
  archivePrefix = {arXiv},
  primaryClass  = {quant-ph}
}

@article{Mo_Settembrini2022VacuumCorrelationsNatCommun,
  title   = {Detection of quantum-vacuum field correlations outside the light cone},
  author  = {Settembrini, Francesca Fabiana and Lindel, Frieder and Herter, Alexa Marina and Buhmann, Stefan Yoshi and Faist, J{\'e}r{\^o}me},
  journal = {Nature Communications},
  volume  = {13},
  number  = {1},
  pages   = {3383},
  year    = {2022},
  doi     = {10.1038/s41467-022-31081-1},
  url     = {https://doi.org/10.1038/s41467-022-31081-1}
}

@article{
MO_Endres2016OneDimensionalTweezerArray,
author = {Manuel Endres  and Hannes Bernien  and Alexander Keesling  and Harry Levine  and Eric R. Anschuetz  and Alexandre Krajenbrink  and Crystal Senko  and Vladan Vuletic  and Markus Greiner  and Mikhail D. Lukin },
title = {Atom-by-atom assembly of defect-free one-dimensional cold atom arrays},
journal = {Science},
volume = {354},
number = {6315},
pages = {1024-1027},
year = {2016},
doi = {10.1126/science.aah3752},
}

@article{MO_Kaufman2021QuantumScienceTweezers,
author = {Kaufman, Adam M. and Ni, Kang-Kuen},
title = {Quantum science with optical tweezer arrays of ultracold atoms and molecules},
journal = {Nature Physics},
volume = {17},
pages = {1324--1333},
year = {2021},
doi = {10.1038/s41567-021-01357-2}
}

@article{Mo_Grusdt2025RepProgPhys_BosePolaronReview,
  title   = {Impurities and polarons in bosonic quantum gases: a review on recent progress},
  author  = {Grusdt, Fabian and Mostaan, Nader and Demler, Eugene and Pe{\~n}a Ardila, Luis A.},
  journal = {Reports on Progress in Physics},
  volume  = {88},
  number  = {6},
  pages   = {066401},
  year    = {2025},
  doi     = {10.1088/1361-6633/add94b}
}

@misc{MO_Massignan2025PolaronsReview,
  title         = {Polarons in atomic gases and two-dimensional semiconductors},
  author        = {Massignan, Pietro and Schmidt, Richard and Astrakharchik, Grigori E. and {\.I}mamo{\u{g}}lu, Ata{\c{c}} and Zwierlein, Martin and Arlt, Jan J. and Bruun, Georg M.},
  year          = {2025},
  eprint        = {2501.09618},
  archivePrefix = {arXiv},
  primaryClass  = {cond-mat.quant-gas},
  doi           = {10.48550/arXiv.2501.09618}
}

@article{MO_Garay2000SonicAnalogPRL,
  title   = {Sonic Analog of Gravitational Black Holes in {Bose-Einstein} Condensates},
  author  = {Garay, L. J. and Anglin, J. R. and Cirac, J. I. and Zoller, P.},
  journal = {Physical Review Letters},
  volume  = {85},
  number  = {22},
  pages   = {4643--4647},
  year    = {2000},
  doi     = {10.1103/PhysRevLett.85.4643}
}

@book{Mo_PitaevskiiStringari2018ReprintBECSuperfluidity,
  title     = {Bose--Einstein Condensation and Superfluidity},
  author    = {Pitaevskii, Lev P. and Stringari, Sandro},
  publisher = {Oxford University Press},
  address   = {Oxford},
  year      = {2018},
  series    = {International Series of Monographs on Physics},
  volume    = {164},
  isbn      = {9780198824435},
  note      = {Reprint}
}

@article{Mo_TolosaSimeon2022CurvedExpandingBEC,
  title   = {Curved and expanding spacetime geometries in {B}ose--{E}instein condensates},
  author  = {Tolosa-Sime{\'o}n, Mireia and Parra-L{\'o}pez, {\'A}lvaro and S{\'a}nchez-Kuntz, Natalia and Haas, Tobias and Viermann, Celia and Sparn, Marius and Liebster, Nikolas and Hans, Maurus and Kath, Elinor and Strobel, Helmut and Oberthaler, Markus K. and Floerchinger, Stefan},
  journal = {Physical Review A},
  volume  = {106},
  number  = {3},
  pages   = {033313},
  year    = {2022},
  doi     = {10.1103/PhysRevA.106.033313},
  eprint  = {2202.10441},
  archivePrefix = {arXiv},
  primaryClass  = {cond-mat.quant-gas}
}

@article{MO_Scelle2013MotionalCoherence,
  title   = {Motional Coherence of Fermions Immersed in a {Bose gas}},
  author  = {Scelle, R. and Rentrop, T. and Trautmann, A. and Schuster, T. and Oberthaler, M. K.},
  journal = {Physical Review Letters},
  volume  = {111},
  number  = {7},
  pages   = {070401},
  year    = {2013},
  doi     = {10.1103/PhysRevLett.111.070401},
  eprint  = {1306.3308},
  archivePrefix = {arXiv},
  primaryClass  = {cond-mat.quant-gas}
}

@article{Mo_Streed2006LargeAtomNumberBEC,
  title   = {Large atom number {Bose-Einstein} condensate machines},
  author  = {Streed, E. W. and Schabel, S. and Mun, J. and Boyd, M. and Campbell, G. K. and Ketterle, W. and Pritchard, D. E.},
  journal = {Review of Scientific Instruments},
  volume  = {77},
  number  = {2},
  pages   = {023106},
  year    = {2006},
  doi     = {10.1063/1.2163977}
}

@article{MO_lattices_RevModPhys,
  title = {Dynamics of {Bose-Einstein} condensates in optical lattices},
  author = {Morsch, Oliver and Oberthaler, Markus},
  journal = {Rev. Mod. Phys.},
  volume = {78},
  issue = {1},
  pages = {179--215},
  numpages = {0},
  year = {2006},
  month = {Feb},
  publisher = {American Physical Society},
  doi = {10.1103/RevModPhys.78.179},
  url = {https://link.aps.org/doi/10.1103/RevModPhys.78.179}
}

@article{MO_tuneoutwavelength,
  title = {Precision measurement of the $^{87}\text{Rb}$ tune-out wavelength in the hyperfine ground state $F=1$ at 790 nm},
  author = {Schmidt, Felix and Mayer, Daniel and Hohmann, Michael and Lausch, Tobias and Kindermann, Farina and Widera, Artur},
  journal = {Phys. Rev. A},
  volume = {93},
  issue = {2},
  pages = {022507},
  numpages = {7},
  year = {2016},
  month = {Feb},
  publisher = {American Physical Society},
  doi = {10.1103/PhysRevA.93.022507},
  url = {https://link.aps.org/doi/10.1103/PhysRevA.93.022507}
}

@article{MO_KRbfeshbach_first,
  title = {Feshbach spectroscopy of a $\mathrm{K}\text{\ensuremath{-}}\mathrm{Rb}$ atomic mixture},
  author = {Ferlaino, Francesca and D'Errico, Chiara and Roati, Giacomo and Zaccanti, Matteo and Inguscio, Massimo and Modugno, Giovanni and Simoni, Andrea},
  journal = {Phys. Rev. A},
  volume = {73},
  issue = {4},
  pages = {040702},
  numpages = {4},
  year = {2006},
  month = {Apr},
  publisher = {American Physical Society},
  doi = {10.1103/PhysRevA.73.040702},
  url = {https://link.aps.org/doi/10.1103/PhysRevA.73.040702}
}

@article{MO_KRbFeshbach_2008,
  title = {Near-threshold model for ultracold {KRb} dimers from interisotope Feshbach spectroscopy},
  author = {Simoni, Andrea and Zaccanti, Matteo and D'Errico, Chiara and Fattori, Marco and Roati, Giacomo and Inguscio, Massimo and Modugno, Giovanni},
  journal = {Phys. Rev. A},
  volume = {77},
  issue = {5},
  pages = {052705},
  numpages = {8},
  year = {2008},
  month = {May},
  publisher = {American Physical Society},
  doi = {10.1103/PhysRevA.77.052705},
  url = {https://link.aps.org/doi/10.1103/PhysRevA.77.052705}
}

@article{MO_Unruhdetector_with_lattice_Reznik,
  title = {Methods for Detecting Acceleration Radiation in a {Bose-Einstein} Condensate},
  author = {Retzker, A. and Cirac, J. I. and Plenio, M. B. and Reznik, B.},
  journal = {Phys. Rev. Lett.},
  volume = {101},
  issue = {11},
  pages = {110402},
  numpages = {4},
  year = {2008},
  month = {Sep},
  publisher = {American Physical Society},
  doi = {10.1103/PhysRevLett.101.110402},
  url = {https://link.aps.org/doi/10.1103/PhysRevLett.101.110402}
}

@article{MO_CasimirIacopo,
  title = {Casimir Forces and Quantum Friction from Ginzburg Radiation in Atomic {Bose-Einstein} Condensates},
  author = {Marino, Jamir and Recati, Alessio and Carusotto, Iacopo},
  journal = {Phys. Rev. Lett.},
  volume = {118},
  issue = {4},
  pages = {045301},
  numpages = {6},
  year = {2017},
  month = {Jan},
  publisher = {American Physical Society},
  doi = {10.1103/PhysRevLett.118.045301},
  url = {https://link.aps.org/doi/10.1103/PhysRevLett.118.045301}
}

@article{MO_zero_crossing_KRb,
  title = {Double Species {Bose-Einstein} Condensate with Tunable Interspecies Interactions},
  author = {Thalhammer, G. and Barontini, G. and De Sarlo, L. and Catani, J. and Minardi, F. and Inguscio, M.},
  journal = {Phys. Rev. Lett.},
  volume = {100},
  issue = {21},
  pages = {210402},
  numpages = {4},
  year = {2008},
  month = {May},
  publisher = {American Physical Society},
  doi = {10.1103/PhysRevLett.100.210402},
  url = {https://link.aps.org/doi/10.1103/PhysRevLett.100.210402}
}

@article{MO_KRbfeshbach_precise,
  title = {Tunable dual-species {Bose-Einstein} condensates of $^{39}\mathrm{K}$ and $^{87}\mathrm{Rb}$},
  author = {Wacker, L. and J\o{}rgensen, N. B. and Birkmose, D. and Horchani, R. and Ertmer, W. and Klempt, C. and Winter, N. and Sherson, J. and Arlt, J. J.},
  journal = {Phys. Rev. A},
  volume = {92},
  issue = {5},
  pages = {053602},
  numpages = {9},
  year = {2015},
  month = {Nov},
  publisher = {American Physical Society},
  doi = {10.1103/PhysRevA.92.053602},
  url = {https://link.aps.org/doi/10.1103/PhysRevA.92.053602}
}

@article{MO_RbK_loss,
  title = {Universal Three-Body Physics in Ultracold {KRb} Mixtures},
  author = {Wacker, L. J. and J\o{}rgensen, N. B. and Birkmose, D. and Winter, N. and Mikkelsen, M. and Sherson, J. and Zinner, N. and Arlt, J. J.},
  journal = {Phys. Rev. Lett.},
  volume = {117},
  issue = {16},
  pages = {163201},
  numpages = {6},
  year = {2016},
  month = {Oct},
  publisher = {American Physical Society},
  doi = {10.1103/PhysRevLett.117.163201},
  url = {https://link.aps.org/doi/10.1103/PhysRevLett.117.163201}
}

@article{MO_RbK_loss_remeasured,
  title = {Isotopic Shift of Atom-Dimer Efimov Resonances in K-Rb Mixtures: Critical Effect of Multichannel Feshbach Physics},
  author = {Kato, K. and Wang, Yujun and Kobayashi, J. and Julienne, P. S. and Inouye, S.},
  journal = {Phys. Rev. Lett.},
  volume = {118},
  issue = {16},
  pages = {163401},
  numpages = {5},
  year = {2017},
  month = {Apr},
  publisher = {American Physical Society},
  doi = {10.1103/PhysRevLett.118.163401},
  url = {https://link.aps.org/doi/10.1103/PhysRevLett.118.163401}
}

@article{MO_phononic_lamb_shift,
  title = {Observation of the Phononic Lamb Shift with a Synthetic Vacuum},
  author = {Rentrop, T. and Trautmann, A. and Olivares, F. A. and Jendrzejewski, F. and Komnik, A. and Oberthaler, M. K.},
  journal = {Phys. Rev. X},
  volume = {6},
  issue = {4},
  pages = {041041},
  numpages = {10},
  year = {2016},
  month = {Nov},
  publisher = {American Physical Society},
  doi = {10.1103/PhysRevX.6.041041},
  url = {https://link.aps.org/doi/10.1103/PhysRevX.6.041041}
}

@article{MO_Strobel2014FisherNonGaussianSpin,
  title   = {Fisher information and entanglement of non-{Gaussian} spin states},
  author  = {Strobel, Helmut and Muessel, Wolf and Linnemann, Daniel and Zibold, Tilman and Hume, David B. and Pezz{\'e}, Luca and Smerzi, Augusto and Oberthaler, Markus K.},
  journal = {Science},
  volume  = {345},
  number  = {6195},
  pages   = {424--427},
  year    = {2014},
  doi     = {10.1126/science.1250147}
}

@article{MO_Feshbach_RevModPhys.82.1225,
  title = {Feshbach resonances in ultracold gases},
  author = {Chin, Cheng and Grimm, Rudolf and Julienne, Paul and Tiesinga, Eite},
  journal = {Rev. Mod. Phys.},
  volume = {82},
  issue = {2},
  pages = {1225--1286},
  numpages = {0},
  year = {2010},
  month = {Apr},
  publisher = {American Physical Society},
  doi = {10.1103/RevModPhys.82.1225},
  url = {https://link.aps.org/doi/10.1103/RevModPhys.82.1225}
}

@article{MO_optomechanics_metric,
author = {F. Bemani and R. Roknizadeh and M. H. Naderi},
journal = {J. Opt. Soc. Am. B},
number = {12},
pages = {2519--2527},
publisher = {Optica Publishing Group},
title = {Analog curved spacetimes in the reversed dissipation regime of cavity optomechanics},
volume = {34},
month = {Dec},
year = {2017},
doi = {10.1364/JOSAB.34.002519},
}

@misc{sergi,
      title={Bipartite entanglement harvesting with multiple detectors}, 
      author={Santeri Salomaa and Esko Keski-Vakkuri and Sergi Nadal-Gisbert},
      year={2026},
      eprint={2604.13869},
      archivePrefix={arXiv},
      primaryClass={quant-ph},
      url={https://arxiv.org/abs/2604.13869}, 
}

@misc{marcos,
      title={Optimization of entanglement harvesting with arbitrary temporal profiles: the limit of second order perturbation theory}, 
      author={Marcos Morote-Balboa and T. Rick Perche},
      year={2026},
      eprint={2604.06303},
      archivePrefix={arXiv},
      primaryClass={quant-ph},
      url={https://arxiv.org/abs/2604.06303}, 
}

@article{MO_Falque2025PolaritonCurvedSpacetimes,
author = {Falque, K{\'e}vin and Delhom, Adri{\`a} and Glorieux, Quentin and Giacobino, Elisabeth and Bramati, Alberto and Jacquet, Maxime J.},
title = {Polariton Fluids as Quantum Field Theory Simulators on Tailored Curved Spacetimes},
journal = {Physical Review Letters},
volume = {135},
pages = {023401},
year = {2025},
doi = {10.1103/t5dh-rx6w}
}

@article{MO_acousticquantumvacuum,
  title = {Unconventional cavity optomechanics: Nonlinear control of phonons in the acoustic quantum vacuum},
  author = {Wang, Xin and Qin, Wei and Miranowicz, Adam and Savasta, Salvatore and Nori, Franco},
  journal = {Phys. Rev. A},
  volume = {100},
  issue = {6},
  pages = {063827},
  numpages = {13},
  year = {2019},
  month = {Dec},
  publisher = {American Physical Society},
  doi = {10.1103/PhysRevA.100.063827},
  url = {https://link.aps.org/doi/10.1103/PhysRevA.100.063827}
}

@article{generalPD,
  title = {Localized nonrelativistic quantum systems in curved spacetimes: A general characterization of particle detector models},
  author = {Perche, T. Rick},
  journal = {Phys. Rev. D},
  volume = {106},
  issue = {2},
  pages = {025018},
  numpages = {20},
  year = {2022},
  month = {Jul},
  publisher = {American Physical Society},
  doi = {10.1103/PhysRevD.106.025018},
  url = {https://link.aps.org/doi/10.1103/PhysRevD.106.025018}
}

@article{Unruh1976,
  title = {Notes on black-hole evaporation},
  author = {Unruh, W. G.},
  journal = {Phys. Rev. D},
  volume = {14},
  issue = {4},
  pages = {870--892},
  numpages = {0},
  year = {1976},
  month = {Aug},
  publisher = {American Physical Society},
  doi = {10.1103/PhysRevD.14.870},
  url = {https://link.aps.org/doi/10.1103/PhysRevD.14.870}
}

@book{DeWitt,
	Address = {Cambridge, UK},
	Author = {B. DeWitt},
	Publisher = {Cambridge University Press},
	Title = {General Relativity; an Einstein Centenary Survey},
	Year = {1980}}

@article{us,
  title = {General relativistic quantum optics: Finite-size particle detector models in curved spacetimes},
  author = {Mart\'{\i}n-Mart\'{\i}nez, Eduardo and Perche, T. Rick and de S. L. Torres, Bruno},
  journal = {Phys. Rev. D},
  volume = {101},
  issue = {4},
  pages = {045017},
  numpages = {10},
  year = {2020},
  month = {Feb},
  publisher = {American Physical Society},
  doi = {10.1103/PhysRevD.101.045017},
  url = {https://link.aps.org/doi/10.1103/PhysRevD.101.045017}
}

@article{mine,
  title = {General features of the thermalization of particle detectors and the {U}nruh effect},
  author = {Perche, T. Rick},
  journal = {Phys. Rev. D},
  volume = {104},
  issue = {6},
  pages = {065001},
  numpages = {14},
  year = {2021},
  month = {Sep},
  publisher = {American Physical Society},
  doi = {10.1103/PhysRevD.104.065001},
  url = {https://link.aps.org/doi/10.1103/PhysRevD.104.065001}
}

@article{FewsterVerch,
  title = {Quantum {F}ields and {L}ocal {M}easurements},
  author = {Fewster, Christopher J. and Verch, Rainer},
  journal = {Commun. Math. Phys.},
  volume = {378},
  pages = {851--889},
  year = {2020},
  month = {},
  publisher = {},
  doi = {10.1007/s00220-020-03800-6},
  url = {https://link.springer.com/article/10.1007/s00220-020-03800-6}
}

@article{UwePRD2004,
  title = {Observer dependence for the phonon content of the sound field living on the effective curved space-time background of a {Bose-Einstein} condensate},
  author = {Fedichev, Petr O. and Fischer, Uwe R.},
  journal = {Phys. Rev. D},
  volume = {69},
  issue = {6},
  pages = {064021},
  numpages = {9},
  year = {2004},
  month = {Mar},
  publisher = {American Physical Society},
  doi = {10.1103/PhysRevD.69.064021},
  url = {https://link.aps.org/doi/10.1103/PhysRevD.69.064021}
}

@misc{fewster3,
      title={Measurement in Quantum Field Theory}, 
      author={Christopher J. Fewster and Rainer Verch},
      year={2023},
      eprint={2304.13356},
      archivePrefix={arXiv},
      primaryClass={math-ph}
}

@misc{QFTPD,
      title={{Particle Detectors from Localized Quantum Field Theories}}, 
      author={T. Rick Perche and Jos\'e Polo-G\'omez and Bruno de S. L. Torres and Eduardo Mart\'in-Mart\'inez},
      year={2023},
      eprint={2308.11698},
      archivePrefix={arXiv},
      primaryClass={quant-ph}
}

@article{Jonsson2,
  title = {Information Transmission Without Energy Exchange},
  author = {Jonsson, Robert H. and Mart\'{i}n-Mart\'{i}nez, Eduardo and Kempf, Achim},
  journal = {Phys. Rev. Lett.},
  volume = {114},
  issue = {11},
  pages = {110505},
  numpages = {5},
  year = {2015},
  month = {Mar},
  publisher = {American Physical Society},
  doi = {10.1103/PhysRevLett.114.110505},
  url = {https://link.aps.org/doi/10.1103/PhysRevLett.114.110505}
}

@article{Nicho1,
  title = {{$\hat{\bm p}\cdot\hat{\bm{A}}$} vs {$\hat{\bm x}\cdot\hat{\bm{E}}$}: Gauge invariance in quantum optics and quantum field theory},
  author = {Funai, Nicholas and Louko, Jorma and Mart\'{\i}n-Mart\'{\i}nez, Eduardo},
  journal = {Phys. Rev. D},
  volume = {99},
  issue = {6},
  pages = {065014},
  numpages = {19},
  year = {2019},
  month = {Mar},
  publisher = {American Physical Society},
  doi = {10.1103/PhysRevD.99.065014},
  url = {https://link.aps.org/doi/10.1103/PhysRevD.99.065014}
}

@article{VidalNegativity,
  title = {Computable measure of entanglement},
  author = {Vidal, G. and Werner, R. F.},
  journal = {Phys. Rev. A},
  volume = {65},
  issue = {3},
  pages = {032314},
  numpages = {11},
  year = {2002},
  month = {Feb},
  publisher = {American Physical Society},
  doi = {10.1103/PhysRevA.65.032314},
  url = {https://link.aps.org/doi/10.1103/PhysRevA.65.032314}
}

@article{ericksonNew,
  title = {When entanglement harvesting is not really harvesting},
  author = {Tjoa, Erickson and Mart\'{\i}n-Mart\'{\i}nez, Eduardo},
  journal = {Phys. Rev. D},
  volume = {104},
  issue = {12},
  pages = {125005},
  numpages = {21},
  year = {2021},
  month = {Dec},
  publisher = {American Physical Society},
  doi = {10.1103/PhysRevD.104.125005},
  url = {https://link.aps.org/doi/10.1103/PhysRevD.104.125005}
}

@article{boris,
  title = {Harvesting entanglement from the gravitational vacuum},
  author = {Perche, T. Rick and Ragula, Boris and Mart\'{\i}n-Mart\'{\i}nez, Eduardo},
  journal = {Phys. Rev. D},
  volume = {108},
  issue = {8},
  pages = {085025},
  numpages = {58},
  year = {2023},
  month = {Oct},
  publisher = {American Physical Society},
  doi = {10.1103/PhysRevD.108.085025},
  url = {https://link.aps.org/doi/10.1103/PhysRevD.108.085025}
}

@inbook{advancesAQFT2015,
    author = "Fredenhagen, Klaus",
    editor = "Brunetti, Romeo and Dappiaggi, Claudio and Fredenhagen, Klaus and Yngvason, Jakob",
    title = "{An Introduction to Algebraic Quantum Field Theory}",
    booktitle = "{Advances in algebraic quantum field theory}",
    doi = "10.1007/978-3-319-21353-8_1",
    pages = "1--30",
    year = "2015"
}

@inbook{kasiaFewsterIntro,
title = "Algebraic Quantum Field Theory: An introduction",
author = "Fewster, {Christopher John} and Rejzner, {Katarzyna Anna}",
year = "2020",
language = "English",
isbn = "978-3-030-38940-6",
editor = "Felix Finster and Domenico Giulini and Johannes Kleiner and J{\"u}rgen Tolksdorf",
booktitle = "Progress and Visions in Quantum Theory in View of Gravity",
publisher = "Birkhauser",
}

@article{Pozas2016,
  title = {Entanglement harvesting from the electromagnetic vacuum with hydrogenlike atoms},
  author = {Pozas-Kerstjens, Alejandro and Mart\'{i}n-Mart\'{i}nez, Eduardo},
  journal = {Phys. Rev. D},
  volume = {94},
  issue = {6},
  pages = {064074},
  numpages = {27},
  year = {2016},
  month = {Sep},
  publisher = {American Physical Society},
  doi = {10.1103/PhysRevD.94.064074},
  url = {https://link.aps.org/doi/10.1103/PhysRevD.94.064074}
}

@article{Pozas-Kerstjens:2015,
	Author = {Pozas-Kerstjens, Alejandro and Mart\'{i}n-Mart\'{i}nez, Eduardo},
	Doi = {10.1103/PhysRevD.92.064042},
	Issue = {6},
	Journal = {Phys. Rev. D},
	Month = {Sep},
	Numpages = {18},
	Pages = {064042},
	Publisher = {American Physical Society},
	Title = {Harvesting correlations from the quantum vacuum},
	Url = {http://link.aps.org/doi/10.1103/PhysRevD.92.064042},
	Volume = {92},
	Year = {2015}}

@article{Reznik1,
	Author = {Benni Reznik and Alex Retzker and Jonathan Silman},
	Eid = {042104},
	Journal = {Phys. Rev. A},
	Number = {4},
	Numpages = {4},
	Pages = {042104},
	Publisher = {APS},
	Title = {{Violating Bell's inequalities in vacuum}},
	Url = {http://link.aps.org/abstract/PRA/v71/e042104},
	Volume = {71},
	Year = {2005}}

@article{Valentini1991,
	Author = {Antony Valentini},
	Doi = {http://dx.doi.org/10.1016/0375-9601(91)90952-5},
	Issn = {0375-9601},
	Journal = {Phys. Lett. A},
	Number = {6-7},
	Pages = {321 - 325},
	Title = {Non-local correlations in quantum electrodynamics},
	Url = {http://www.sciencedirect.com/science/article/pii/0375960191909525},
	Volume = {153},
	Year = {1991}}

@article{Reznik2003,
	Author = {Reznik, Benni},
	Doi = {10.1023/A:1022875910744},
	Issn = {0015-9018},
	Journal = {Foundations of Physics},
	Language = {English},
	Number = {1},
	Pages = {167-176},
	Publisher = {Kluwer Academic Publishers-Plenum Publishers},
	Title = {Entanglement from the Vacuum},
	Url = {http://dx.doi.org/10.1023/A%3A1022875910744},
	Volume = {33},
	Year = {2003}}

@article{witten,
  title = {APS Medal for Exceptional Achievement in Research: Invited article on entanglement properties of quantum field theory},
  author = {Witten, Edward},
  journal = {Rev. Mod. Phys.},
  volume = {90},
  issue = {4},
  pages = {045003},
  numpages = {38},
  year = {2018},
  month = {Oct},
  publisher = {American Physical Society},
  doi = {10.1103/RevModPhys.90.045003},
  url = {https://link.aps.org/doi/10.1103/RevModPhys.90.045003}
}

@book{Haag,
  doi = {10.1007/978-3-642-97306-2},
  url = {https://doi.org/10.1007/978-3-642-97306-2},
  year = {1992},
  publisher = {Springer-Verlag Berlin Heidelberg},
  author = {Rudolf Haag},
  title = {Local Quantum Physics: Fields, Particles, Algebras}
}

@article{matsasUnruh,
  title = {The {U}nruh effect and its applications},
  author = {Crispino, Lu\'{\i}s C. B. and Higuchi, Atsushi and Matsas, George E. A.},
  journal = {Rev. Mod. Phys.},
  volume = {80},
  issue = {3},
  pages = {787--838},
  numpages = {0},
  year = {2008},
  month = {Jul},
  publisher = {American Physical Society},
  doi = {10.1103/RevModPhys.80.787},
  url = {https://link.aps.org/doi/10.1103/RevModPhys.80.787}
}

@article{Unruh-Wald,
	Author = {Unruh, William G. and Wald, Robert M.},
	Doi = {10.1103/PhysRevD.29.1047},
	Issue = {6},
	Journal = {Phys. Rev. D},
	Month = {Mar},
	Pages = {1047--1056},
	Publisher = {American Physical Society},
	Title = {{What happens when an accelerating observer detects a Rindler particle}},
	Volume = {29},
	Year = {1984}
}

@Article{Hawking1975,
author={Hawking, S. W.},
title={Particle creation by black holes},
journal={Comm. Math. Phys.},
year={1975},
month={Aug},
day={01},
volume={43},
number={3},
pages={199-220},
issn={1432-0916},
doi={10.1007/BF02345020},
url={https://doi.org/10.1007/BF02345020}
}

@article{Takagi,
    author = {Takagi, Shin},
    title = "{{Vacuum Noise and Stress Induced by Uniform Acceleration: Hawking-{Unruh effect} in Rindler Manifold of Arbitrary Dimension}}",
    journal = {Prog. Theor. Phys. Supp.},
    volume = {88},
    pages = {1-142},
    year = {1986},
    month = {03},
    issn = {0375-9687},
    doi = {10.1143/PTP.88.1},
    url = {https://doi.org/10.1143/PTP.88.1},
}

@article{UnruhPhilosophers,
	doi = {10.1016/j.shpsb.2011.04.001},
	volume = {42},
	title = {The {Unruh effect} for Philosophers},
	year = {2011},
	publisher = {Elsevier B.V.},
	author = {John Earman},
	pages = {81--97},
	journal = {STUD HIST PHILOS M P},
	number = {2}
}

@article{remi,
  title = {Particle detectors as witnesses for quantum gravity},
  author = {Faure, R\'emi and Perche, T. Rick and Torres, Bruno de S. L.},
  journal = {Phys. Rev. D},
  volume = {101},
  issue = {12},
  pages = {125018},
  numpages = {9},
  year = {2020},
  month = {Jun},
  publisher = {American Physical Society},
  doi = {10.1103/PhysRevD.101.125018},
  url = {https://link.aps.org/doi/10.1103/PhysRevD.101.125018}
}

@article{richard,
  title = {Quantum delocalization, gauge, and quantum optics: Light-matter interaction in relativistic quantum information},
  author = {Lopp, Richard and Mart\'{i}n-Mart\'{i}nez, Eduardo},
  journal = {Phys. Rev. A},
  volume = {103},
  issue = {1},
  pages = {013703},
  numpages = {20},
  year = {2021},
  month = {Jan},
  doi = {10.1103/PhysRevA.103.013703},
  url = {https://link.aps.org/doi/10.1103/PhysRevA.103.013703}
}

@article{teleportation,
  title = {Quantum measurement information as a key to energy extraction from local vacuums},
  author = {Hotta, Masahiro},
  journal = {Phys. Rev. D},
  volume = {78},
  issue = {4},
  pages = {045006},
  numpages = {9},
  year = {2008},
  month = {Aug},
  publisher = {American Physical Society},
  doi = {10.1103/PhysRevD.78.045006},
  url = {https://link.aps.org/doi/10.1103/PhysRevD.78.045006}
}

@misc{Hotta2011,
  title={{Q}uantum {E}nergy {T}eleportation: {A}n {I}ntroductory {R}eview}, 
  author={Masahiro Hotta},
  year={2011},
  eprint={1101.3954},
  archivePrefix={arXiv},
  primaryClass={quant-ph}
}

@article{pitelli,
  title = {Angular momentum based graviton detector},
  author = {Pitelli, J. P. M. and Perche, T. Rick},
  journal = {Phys. Rev. D},
  volume = {104},
  issue = {6},
  pages = {065016},
  numpages = {9},
  year = {2021},
  month = {Sep},
  publisher = {American Physical Society},
  doi = {10.1103/PhysRevD.104.065016},
  url = {https://link.aps.org/doi/10.1103/PhysRevD.104.065016}
}

@article{B,
  title = {Spin Entanglement Witness for Quantum Gravity},
  author = {Bose, Sougato and Mazumdar, Anupam and Morley, Gavin W. and Ulbricht, Hendrik and Toro\ifmmode \check{s}\else \v{s}\fi{}, Marko and Paternostro, Mauro and Geraci, Andrew A. and Barker, Peter F. and Kim, M. S. and Milburn, Gerard},
  journal = {Phys. Rev. Lett.},
  volume = {119},
  issue = {24},
  pages = {240401},
  numpages = {6},
  year = {2017},
  month = {Dec},
  publisher = {American Physical Society},
  doi = {10.1103/PhysRevLett.119.240401},
  url = {https://link.aps.org/doi/10.1103/PhysRevLett.119.240401}
}

@article{AreaLawsBEC2025,
  title = {Area laws and thermalization from classical entropies in a {Bose-Einstein} condensate},
  author = {Deller, Yannick and G\"arttner, Martin and Haas, Tobias and Oberthaler, Markus K. and Reh, Moritz and Strobel, Helmut},
  journal = {Phys. Rev. A},
  volume = {112},
  issue = {1},
  pages = {L011303},
  numpages = {7},
  year = {2025},
  month = {Jul},
  publisher = {American Physical Society},
  doi = {10.1103/7jzy-g3vd},
  url = {https://link.aps.org/doi/10.1103/7jzy-g3vd}
}

@article{KlcoEntStruc2023,
  title = {Entanglement structures in quantum field theories: Negativity cores and bound entanglement in the vacuum},
  author = {Klco, Natalie and Beck, D. H. and Savage, Martin J.},
  journal = {Phys. Rev. A},
  volume = {107},
  issue = {1},
  pages = {012415},
  numpages = {22},
  year = {2023},
  month = {Jan},
  publisher = {American Physical Society},
  doi = {10.1103/PhysRevA.107.012415},
  url = {https://link.aps.org/doi/10.1103/PhysRevA.107.012415}
}

@article{KlcoEntStruc2023II,
  title = {Entanglement structures in quantum field theories. {II.} Distortions of vacuum correlations through the lens of local observers},
  author = {Klco, Natalie and Beck, D. H.},
  journal = {Phys. Rev. A},
  volume = {108},
  issue = {1},
  pages = {012429},
  numpages = {16},
  year = {2023},
  month = {Jul},
  publisher = {American Physical Society},
  doi = {10.1103/PhysRevA.108.012429},
  url = {https://link.aps.org/doi/10.1103/PhysRevA.108.012429}
}

@article{matheusDerivative2023,
  title = {Duality between amplitude and derivative coupled particle detectors in the limit of large energy gaps},
  author = {Perche, T. Rick and Zambianco, Matheus H.},
  journal = {Phys. Rev. D},
  volume = {108},
  issue = {4},
  pages = {045017},
  numpages = {13},
  year = {2023},
  month = {Aug},
  publisher = {American Physical Society},
  doi = {10.1103/PhysRevD.108.045017},
  url = {https://link.aps.org/doi/10.1103/PhysRevD.108.045017}
}

@Article{algebras2023,
author={Chandrasekaran, Venkatesa
and Longo, Roberto
and Penington, Geoff
and Witten, Edward},
title={An algebra of observables for {de Sitter} space},
journal={J. High Energy Phys.},
year={2023},
month={Feb},
day={07},
volume={2023},
number={2},
pages={82},
issn={1029-8479},
doi={10.1007/JHEP02(2023)082},
url={https://doi.org/10.1007/JHEP02(2023)082}
}

@Article{FewsterAlgebras,
author={Fewster, Christopher J.
and Janssen, Daan W.
and Loveridge, Leon Deryck
and Rejzner, Kasia
and Waldron, James},
title={Quantum Reference Frames, Measurement Schemes and the Type of Local Algebras in Quantum Field Theory},
journal={Comm. in Math. Phys.},
year={2024},
month={Dec},
day={18},
volume={406},
number={1},
pages={19},
issn={1432-0916},
doi={10.1007/s00220-024-05180-7},
url={https://doi.org/10.1007/s00220-024-05180-7}
}

@article{JormaHowThermal,
  title = {{Inextendible Schwarzschild black hole with a single exterior: How thermal is the {Hawking} radiation?}},
  author = {Louko, Jorma and Marolf, Donald},
  journal = {Phys. Rev. D},
  volume = {58},
  issue = {2},
  pages = {024007},
  numpages = {27},
  year = {1998},
  month = {Jun},
  publisher = {American Physical Society},
  doi = {10.1103/PhysRevD.58.024007},
  url = {https://link.aps.org/doi/10.1103/PhysRevD.58.024007}
}

@article{JormaHawking,
  title = {{Static detectors and circular-geodesic detectors on the Schwarzschild black hole}},
  author = {Hodgkinson, Lee and Louko, Jorma and Ottewill, Adrian C.},
  journal = {Phys. Rev. D},
  volume = {89},
  issue = {10},
  pages = {104002},
  numpages = {26},
  year = {2014},
  month = {May},
  publisher = {American Physical Society},
  doi = {10.1103/PhysRevD.89.104002},
  url = {https://link.aps.org/doi/10.1103/PhysRevD.89.104002}
}

@article{derivativeJorma,
doi = {10.1088/0264-9381/31/24/245007},
url = {https://dx.doi.org/10.1088/0264-9381/31/24/245007},
year = {2014},
month = {nov},
publisher = {IOP Publishing},
volume = {31},
number = {24},
pages = {245007},
author = {Benito A Juárez-Aubry and Jorma Louko},
title = {{Onset and decay of the 1 + 1 {Hawking}–{Unruh effect}: what the derivative-coupling detector saw}},
journal = {Class. Quantum Gravity}
}

@article{adamDerivative2024,
  title = {Derivative coupling enables genuine entanglement harvesting in causal communication},
  author = {Teixid\'o-Bonfill, Adam and Mart\'{\i}n-Mart\'{\i}nez, Eduardo},
  journal = {Phys. Rev. D},
  volume = {110},
  issue = {10},
  pages = {105016},
  numpages = {8},
  year = {2024},
  month = {Nov},
  publisher = {American Physical Society},
  doi = {10.1103/PhysRevD.110.105016},
  url = {https://link.aps.org/doi/10.1103/PhysRevD.110.105016}
}

@article{quantClass,
  title = {Role of quantum degrees of freedom of relativistic fields in quantum information protocols},
  author = {Perche, T. Rick and Mart\'{\i}n-Mart\'{\i}nez, Eduardo},
  journal = {Phys. Rev. A},
  volume = {107},
  issue = {4},
  pages = {042612},
  numpages = {20},
  year = {2023},
  month = {Apr},
  publisher = {American Physical Society},
  doi = {10.1103/PhysRevA.107.042612},
  url = {https://link.aps.org/doi/10.1103/PhysRevA.107.042612}
}

@article{max,
doi = {10.1088/1361-6382/ac1b08},
url = {https://dx.doi.org/10.1088/1361-6382/ac1b08},
year = {2021},
month = {sep},
publisher = {IOP Publishing},
volume = {38},
number = {19},
pages = {195029},
author = {Maximilian H Ruep},
title = {Weakly coupled local particle detectors cannot harvest entanglement},
journal = {Class. Quantum Gravity}
}

@article{closedform2024,
  title = {Closed-form expressions for smeared bidistributions of a massless scalar field: Nonperturbative and asymptotic results in relativistic quantum information},
  author = {Perche, T. Rick},
  journal = {Phys. Rev. D},
  volume = {110},
  issue = {2},
  pages = {025013},
  numpages = {21},
  year = {2024},
  month = {Jul},
  publisher = {American Physical Society},
  doi = {10.1103/PhysRevD.110.025013},
  url = {https://link.aps.org/doi/10.1103/PhysRevD.110.025013}
}

@article{unruhExperimentalBHE1981,
  title = {Experimental {{Black-Hole Evaporation}}?},
  author = {Unruh, W. G.},
  date = {1981-05-25},
  journal = {Phys. Rev. Lett.},
  volume = {46},
  number = {21},
  pages = {1351--1353},
  publisher = {American Physical Society},
  doi = {10.1103/PhysRevLett.46.1351},
  url = {https://link.aps.org/doi/10.1103/PhysRevLett.46.1351},
  urldate = {2023-03-31}
}

@article{FirstKRb2016,
  title = {Bose Polarons in the Strongly Interacting Regime},
  author = {Hu, Ming-Guang and Van de Graaff, Michael J. and Kedar, Dhruv and Corson, John P. and Cornell, Eric A. and Jin, Deborah S.},
  journal = {Phys. Rev. Lett.},
  volume = {117},
  issue = {5},
  pages = {055301},
  numpages = {6},
  year = {2016},
  month = {Jul},
  publisher = {American Physical Society},
  doi = {10.1103/PhysRevLett.117.055301},
  url = {https://link.aps.org/doi/10.1103/PhysRevLett.117.055301}
}

@article{barceloAnalogueGravity2005,
  title = {Analogue {{Gravity}}},
  author = {Barcelo, Carlos and Liberati, Stefano and Visser, Matt},
  date = {2005-12},
  journal = {Living Rev. Relativ.},
  volume = {8},
  number = {1},
  eprint = {gr-qc/0505065},
  eprinttype = {arXiv},
  pages = {12},
  issn = {2367-3613, 1433-8351},
  doi = {10.12942/lrr-2005-12},
  url = {http://arxiv.org/abs/gr-qc/0505065},
  urldate = {2025-06-13}
}

@article{cucchiettiStrongCouplingPolaronsDilute2006,
  title = {Strong-{{Coupling Polarons}} in {{Dilute Gas {Bose-Einstein} Condensates}}},
  author = {Cucchietti, F. M. and Timmermans, E.},
  date = {2006-06-01},
  journal = {Phys. Rev. Lett.},
  volume = {96},
  number = {21},
  pages = {210401},
  publisher = {American Physical Society},
  doi = {10.1103/PhysRevLett.96.210401},
  url = {https://link.aps.org/doi/10.1103/PhysRevLett.96.210401},
  urldate = {2025-10-22}
}

@article{tempereFeynmanPathintegralTreatment2009,
  title = {Feynman Path-Integral Treatment of the {{BEC-impurity}} Polaron},
  author = {Tempere, J. and Casteels, W. and Oberthaler, M. K. and Knoop, S. and Timmermans, E. and Devreese, J. T.},
  date = {2009-11-10},
  journal = {Phys. Rev. B},
  volume = {80},
  number = {18},
  pages = {184504},
  publisher = {American Physical Society},
  doi = {10.1103/PhysRevB.80.184504},
  url = {https://link.aps.org/doi/10.1103/PhysRevB.80.184504},
  urldate = {2025-07-29}
}

@article{frolich1954,
       author = {{Fr{\"o}hlich}, H.},
        title = "{Electrons in lattice fields}",
      journal = {Advances in Physics},
         year = 1954,
        month = jul,
       volume = {3},
       number = {11},
        pages = {325-361},
          doi = {10.1080/00018735400101213},
       adsurl = {https://ui.adsabs.harvard.edu/abs/1954AdPhy...3..325F}
}

@book{pitaevskii2016bose,
  title={{Bose-Einstein} condensation and superfluidity},
  author={Pitaevskii, Lev and Stringari, Sandro},
  volume={164},
  year={2016},
  publisher={Oxford University Press}
}

@book{peskin2018introduction,
  title={An Introduction to quantum field theory},
  author={Peskin, Michael E},
  year={2018},
  publisher={CRC press}
}

@article{lee1957many,
  title={Many-body problem in quantum mechanics and quantum statistical mechanics},
  author={Lee, TD and Yang, CN},
  journal={Physical Review},
  volume={105},
  number={3},
  pages={1119},
  year={1957},
  publisher={APS}
}

@article{lee1959low,
  title={Low-temperature behavior of a dilute Bose system of hard spheres. {II.} Nonequilibrium properties},
  author={Lee, TD and Yang, CN},
  journal={Physical Review},
  volume={113},
  number={6},
  pages={1406},
  year={1959},
  publisher={APS}
}

@misc{adamExperimentalEH2025,
      title={Towards an experimental implementation of entanglement harvesting in superconducting circuits: effect of detector gap variation on entanglement harvesting}, 
      author={Adam Teixidó-Bonfill and Xi Dai and Adrian Lupascu and Eduardo Martín-Martínez},
      year={2025},
      eprint={2505.01516},
      archivePrefix={arXiv},
      primaryClass={quant-ph},
      url={https://arxiv.org/abs/2505.01516}, 
}

@article{Martin-Martinez_Aasen_Kempf_2013, title={Processing quantum information with relativistic motion of atoms}, volume={110}, ISSN={0031-9007, 1079-7114}, DOI={10.1103/PhysRevLett.110.160501}, number={16}, journal={Phys. Rev. Lett.}, author={Martin-Martinez, Eduardo and Aasen, David and Kempf, Achim}, year={2013}, pages={160501} }

@article{Bruschi_Dragan_Lee_Fuentes_Louko_2013, title={Relativistic Motion Generates Quantum Gates and Entanglement Resonances}, volume={111}, DOI={10.1103/PhysRevLett.111.090504}, number={9}, journal={Phys. Rev. Lett.}, publisher={American Physical Society}, author={Bruschi, David Edward and Dragan, Andrzej and Lee, Antony R. and Fuentes, Ivette and Louko, Jorma}, year={2013}, pages={090504} }

@article{Martín-Martínez_Sutherland_2014, title={Quantum gates via relativistic remote control}, volume={739}, ISSN={0370-2693}, DOI={10.1016/j.physletb.2014.10.038}, journal={Phys. Lett. B}, author={Martín-Martínez, Eduardo and Sutherland, Chris}, year={2014}, pages={74–82} }

@article{Layden_Martin-Martinez_Kempf_2016, title={Universal scheme for indirect quantum control}, volume={93}, ISSN={2469-9926, 2469-9934}, DOI={10.1103/PhysRevA.93.040301}, number={4}, journal={Phys. Rev. A}, author={Layden, David and Martin-Martinez, Eduardo and Kempf, Achim}, year={2016}, pages={040301} }

@article{phil2025,
  title = {Universal Quantum Computer from Relativistic Motion},
  author = {LeMaitre, Philip A. and Perche, T. Rick and Krumm, Marius and Briegel, Hans J.},
  journal = {Phys. Rev. Lett.},
  volume = {134},
  issue = {19},
  pages = {190601},
  numpages = {7},
  year = {2025},
  month = {May},
  publisher = {American Physical Society},
  doi = {10.1103/PhysRevLett.134.190601},
  url = {https://link.aps.org/doi/10.1103/PhysRevLett.134.190601}
}

@misc{borisQET2025,
      title={A review of applications of Quantum Energy Teleportation: from experimental tests to thermodynamics and spacetime engineering}, 
      author={Boris Ragula and Eduardo Martín-Martínez},
      year={2025},
      eprint={2505.04689},
      archivePrefix={arXiv},
      primaryClass={quant-ph},
      url={https://arxiv.org/abs/2505.04689}, 
}

@article{PhysRevLett.91.240407,
  title = {Gibbons-{Hawking} Effect in the Sonic {de Sitter} Space-Time of an Expanding {Bose-Einstein}-Condensed Gas},
  author = {Fedichev, Petr O. and Fischer, Uwe R.},
  journal = {Phys. Rev. Lett.},
  volume = {91},
  issue = {24},
  pages = {240407},
  numpages = {4},
  year = {2003},
  month = {Dec},
  publisher = {American Physical Society},
  doi = {10.1103/PhysRevLett.91.240407},
  url = {https://link.aps.org/doi/10.1103/PhysRevLett.91.240407}
}

@misc{tian2025,
      title={Harvesting entanglement from the {Lorentz}-violating quantum field vacuum in a dipolar {Bose-Einstein} condensate}, 
      author={Zehua Tian and Weiping Yao and Xiaobao Liu and Mengjie Wang and Jieci Wang and Jiliang Jing},
      year={2025},
      eprint={2512.09263},
      archivePrefix={arXiv},
      primaryClass={quant-ph},
      url={https://arxiv.org/abs/2512.09263}, 
}

\appendix

\onecolumngrid



\section{Phonons as relativistic quantum fields}
\label{app:BECQFT}

Here we show that the phase and density fluctuations of a uniform BEC are analogous to relativistic quantum fields in Minkowski spacetime. The (time-independent) field operator for the condensate is
\begin{equation*}
\hat \Psi(\bm x) = \frac{1}{\sqrt{V}} \sum_{\bm k} \hat a_{\bm k} e^{i \bm k \cdot \bm x}
\end{equation*}
in a plane wave basis. Under the Bogoliubov approximation, we can split the field operator as
\begin{equation*}
\hat \Psi(\bm x) = \Psi_0 + \delta\hat \Psi(\bm x) =\sqrt{\frac{N}{V}} + \frac{1}{\sqrt{V}}\sum_{\bm k \neq 0} a_{\bm k} e^{i \bm k \cdot \bm x}
\end{equation*}
where $\delta\hat \Psi(\bm x)$ is the field operator of the fluctuations over the background. In the following, all the sums are meant excluding the $\bm k = 0$ component. In the Madelung representation $\Psi = \sqrt{\rho}\,e^{i\phi}$, expanding to first order fluctuations of the density and phase fields 
\begin{equation}
    \delta\hat\Psi(\bm x) = \frac{1}{2\sqrt{\rho_0}}\delta \hat \rho(\bm x) + i \sqrt{\rho_0}\, \delta \hat \phi(\bm x)
\end{equation}
where $\rho_0 \equiv \Psi_0 = \sqrt{N/V}$ is the uniform background density of the condensate. Then
\begin{align*}
 \delta\hat \rho(\bm x) &= \sqrt{\rho_0}\left(\delta\hat\Psi + \delta\hat\Psi^\dagger\right) = \sqrt{\frac{\rho_0}{V}}\sum_{\bm k}\left( \hat a_{\bm k} e^{i \bm k \cdot \bm x} + \hat a_{\bm k}^\dagger e^{-i \bm k \cdot \bm x}\right) \\
  \delta\hat \phi(\bm x) &= \frac{1}{2i\sqrt{\rho_0}}\left(\delta\hat\Psi - \delta\hat\Psi^\dagger\right) = \frac{-i}{2\sqrt{\rho_0 V}}\sum_{\bm k}\left( \hat a_{\bm k} e^{i \bm k \cdot \bm x} - \hat a_{\bm k}^\dagger e^{-i \bm k \cdot \bm x}\right)
\end{align*}
Now we introduce phononic excitations by Bogoliubov transformations \cite{pitaevskii2016bose}. Let
\begin{equation}
    \hat a_{\bm k} = u_{\bm k} \hat b_{\bm k} + v^*_{-\bm k} \hat b^\dagger_{-\bm k}, \qquad \hat a^\dagger_{\bm k} = u^*_{\bm k} \hat b^\dagger_{\bm k} + v_{-\bm k} \hat b_{-\bm k} \,.
\end{equation}
The density and fluctuations field take the form
\begin{align}\label{eq:nphi_bs}
     \delta\hat \rho(\bm x) &= \sqrt{\frac{\rho_0}{V}}\sum_{\bm k}\left\lbrace (u_{\bm k} + v_{\bm k})\hat b_{\bm k} e^{i \bm k \cdot \bm x} + (u^*_{\bm k} + v^*_{\bm k})\hat b_{\bm k}^\dagger e^{-i \bm k \cdot \bm x}\right\rbrace \\
  \delta\hat \phi(\bm x) &= \frac{-i}{2\sqrt{\rho_0 V}}\sum_{\bm k}\left\lbrace (u_{\bm k} - v_{\bm k})\hat b_{\bm k} e^{i \bm k \cdot \bm x} - (u^*_{\bm k} - v^*_{\bm k})\hat b_{\bm k}^\dagger e^{-i \bm k \cdot \bm x}\right\rbrace \,. 
\end{align}
The $u_{\bm k}, v_{\bm k}$ coefficients can be chosen real as
\begin{equation}
    u_{\bm k} = \sqrt{\frac{\hbar^2 |\bm{k}|^2/2m_b + g_{bb}\,\rho_0}{2\hbar\omega_{\bm k}} + \frac{1}{2}}, \qquad v_{-\bm k} = -\sqrt{\frac{\hbar^2 |\bm{k}|^2/2m_b + g_{bb}\,\rho_0}{2\hbar\omega_{\bm k}} - \frac{1}{2}}
\end{equation}
with Bogoliubov dispersion $\omega_{\bm k} = \sqrt{c_s^2 |\bm k|^2 + (\hbar^2 |\bm k|^2/2m_b)^2}$ and $c_s = \sqrt{g_{bb}\, \rho_0/m_b}$. We are interested in the low energy limit $|\bm k| \ll m_b\,c_s/\hbar$, in which the phononic excitations can be considered as a relativistic quantum field. To see this, in this limit one finds%
\begin{align*}
 u_{\bm k} + v_{\bm k} &\approx \sqrt{\frac{g_{bb}\,\rho_0}{2\hbar\omega_{\bm k}} + \frac{1}{2}} - \sqrt{\frac{g_{bb}\,\rho_0}{2\hbar\omega_{\bm k}} - \frac{1}{2}} \approx  \sqrt{\frac{\hbar\omega_{\bm k}}{2g_{bb}\,\rho_0}}\\
 u_{\bm k} - v_{\bm k} &\approx \sqrt{\frac{g_{bb}\,\rho_0}{2\hbar\omega_{\bm k}} + \frac{1}{2}} + \sqrt{\frac{g_{bb}\,\rho_0}{2\hbar\omega_{\bm k}} - \frac{1}{2}} \approx  2\sqrt{\frac{g_{bb}\,\rho_0}{2\hbar\omega_{\bm k}}} \,.
\end{align*}
Substitution in \eqref{eq:nphi_bs} brings
\begin{align}
 \delta\hat \rho(\bm x) &= \frac{1}{\sqrt{V}}\sum_{\bm k}\sqrt{\frac{\hbar\omega_{\bm k}}{2g_{bb}}}\left( \hat b_{\bm k} e^{i \bm k \cdot \bm x} + \hat b_{\bm k}^\dagger e^{-i \bm k \cdot \bm x}\right) \\
  \delta\hat \phi(\bm x) &= \frac{-i}{\sqrt{V}}\sum_{\bm k}\sqrt{\frac{g_{bb}}{2\hbar\omega_{\bm k}}}\left( \hat b_{\bm k} e^{i \bm k \cdot \bm x} - \hat b_{\bm k}^\dagger e^{-i \bm k \cdot \bm x}\right) \,. 
\end{align}
This shows that the quantum fluctuations of density and phase behave as massless relativistic quantum fields in the low energy regime, with dispersion (energy-momentum) relation $\omega_{\bm k} = c_s|\bm k|$, by straight comparison with the analogous expressions for a free massless scalar field in Minkowski spacetime \cite{peskin2018introduction}. The density field operator can be interpreted as the momentum conjugate to the phase field operator \footnote{The signs are not the usual ones because the primary field (phase) is really the imaginary part of the original field (the linear fluctuations of the complex atomic field $\Psi$).}.
The classical Hamiltonian for the low energy theory is
\begin{equation}
H = \int {\rm d}^n x\,\left\lbrace \frac{1}{2g_{bb}} (\delta \rho)^2 - \frac{\rho_0}{2m_b}(\partial_i\delta \phi)(\partial^i\delta \phi)\right\rbrace \,.
\end{equation}
It is straightforward to verify by substitution that the quantum Hamiltonian is
\begin{equation}
\hat H = \sum_{\bm k} \hbar \omega_{\bm k} \left( \hat b_{\bm k}^\dagger \hat b_{\bm k} + \tfrac{1}{2} \right) = \sum_{\bm k} \hbar \omega_{\bm k}\, \hat b_{\bm k}^\dagger \hat b_{\bm k} + E_0
\end{equation}
corresponding to the energy of free phonons plus zero-point energy. The apparently divergent zero-point energy is ``renormalized'' in this simple theory by going to higher order in the interaction expansion \cite{lee1957many, lee1959low}.
We can turn the density and phase fluctuations field operators into time-dependent relativistic form by going from the Schr\"odinger to the Heisenberg picture. Using $e^{i\hat Ht/\hbar} b_{\bm k}e^{-i\hat Ht/\hbar} = b_{\bm k}e^{-i\omega_{\bm k}t}$ and its complex conjugate we obtain
\begin{align}
 \delta\hat \rho(\mf x) &= \frac{1}{\sqrt{V}}\sum_{\bm k}\sqrt{\frac{\hbar\omega_{\bm k}}{2g_{bb}}}\left( \hat b_{\bm k} e^{i \mf k \cdot  \mf x} + \hat b_{\bm k}^\dagger e^{-i \mf k\cdot \mf x}\right) \\
  \delta\hat \phi(\mf x) &= \frac{-i}{\sqrt{V}}\sum_{\bm k}\sqrt{\frac{g_{bb}}{2\hbar\omega_{\bm k}}}\left( \hat b_{\bm k} e^{i \mf k\cdot \mf x} - \hat b_{\bm k}^\dagger e^{-i \mf k \cdot \mf x}\right) \,. 
\end{align}
with $\mf x = (t, \bm x)$ and $\mf k = (\omega_{\bm k}, \bm k)$.
The fields satisfy the usual equal-time commutation relations and
\begin{equation}
     \hbar\frac{\partial}{\partial t} \delta \hat \phi(\mf x) = -g_{bb}\, \delta \hat \rho(\mf x)
\end{equation}
which is the usual relation between density and phase fluctuations in a weakly interacting BEC. 

\section{Calculation of the excitation probability}\label{app:excprob}

The transition amplitude $d$ between times $t_i$ and $t_f$ from the state $\ket{g, 0}$ (impurity in its ground state, no phonons) to the state $\ket{e, \zeta}$ (impurity in excited state, arbitrary number of phonons) for small coupling $\gIB$ is given by first order perturbation theory
\begin{equation}
    d = \frac{i}{\hbar} \int_{t_i}^{t_f} \dd t\; \mel{\zeta, e}{\hat H_I(t)}{g, 0}
\end{equation}
where $\hat H_I(t)$, the interaction term in the Hamiltonian in the interaction picture, is
\begin{equation}
    \hat{H}_I(t) =  \int \dd^3 \bm{x}\,g_{ab}(t) \hat \rho_a(\mf x) \hat \rho_b(\mf x) \,.
\end{equation}
We write $\hat{\rho}_b(\mf x) = \rho_0 + \delta\hat{\rho}(\mf x)$, and drop the classical term $\rho_0$. The impurity density operator in our setting becomes
\begin{equation}
    \hat\rho_a(\mf x) = f^*(\bm{x})e^{i\Omega t} \hat{\sigma}^+ + f(\bm{x})e^{-i\Omega t} \hat{\sigma}^-
\end{equation}
with $f(\bm x)=(\psi_g^*\psi_e)(\bm x)$ in terms of the eigenfunction decomposition of the impurity. The excitation probability $\ket{g} \to \ket{e}$ is the sum over all possible states of the phonon field of the modulus of the transition amplitude squared. Let $\ket{n_{\bm{k}}}$ be a multi-phonon state in Fock space. We have
\begin{align*}
    \mathcal{L} &= \sum_{n_{\bm{k}}} \bigg\vert -\frac{i}{\hbar}   \int_{t_i}^{t_f} \dd t\; \mel{n_{\bm{k}}, e}{H_I(t)}{g, 0} \bigg\vert^2 \\
    &= \frac{1}{\hbar^2}  \int_{t_i}^{t_f}  \int_{t_i}^{t_f} \dd t\, \dd t'\; \gIB(t) \gIB(t')  \\
    &\;\times \int \dd^3 \bm{x}\int \dd^3 \bm{x}'\;  \mel{g}{\hat\rho_a(t', \bm{x}')}{e} \mel{e}{\hat\rho_a(t, \bm{x})}{g}  \times \sum_{n_{\bm{k}}} \mel{0}{\delta\hat \rho(t', \bm{x}')}{n_{\bm{k}}}\mel{n_{\bm{k}}}{\delta\hat \rho(t, \bm{x})}{0} \\
    &= \frac{1}{\hbar^2}  \int_{t_i}^{t_f}  \int_{t_i}^{t_f} \dd t\, \dd t'\; \gIB(t) \gIB(t')\int \dd^3 \bm{x}\int \dd^3 \bm{x}'\; f(\bm{x}')f^*(\bm{x})\mel{0}{\delta\hat{\rho}(t',\bm x')\delta\hat{\rho}(t,\bm x)}{0}e^{i\Omega(t-t')}
    \\
    &= \frac{\bar g_{ab}^2}{\hbar^2}  \int_{t_i}^{t_f}  \int_{t_i}^{t_f} \dd t\, \dd t'\int \dd^3 \bm{x}\int \dd^3 \bm{x}'\; \chi(t) f(\bm{x})\chi(t')f^*(\bm{x}')\mel{0}{\delta\hat{\rho}(\mf{x})\delta\hat{\rho}(\mf{x}')}{0}e^{-i\Omega(t-t')}
\end{align*}
using the properties of the ladder operators and the completeness relation $\mathbf{1} = \sum_{n_{\bm{k}}} \ket{n_{\bm{k}}}\bra{n_{\bm{k}}}$. To get the last equality we wrote $g_{ab}(t) = \bar{g}_{ab}\chi(t)$ with $0 \leq \chi(t)\leq 1$ and performed a change of variables $\mf x \leftrightarrow \mf x'$. Notice that if $\chi(t)$ is only non-zero in a interval $I$ contained in $[t_i,t_f]$, we can instead rewrite the integral as an integration over all of spacetime:
\begin{align}
    \mathcal{L} &= \frac{\bar{g}_{ab}^2}{\hbar^2}  \int\dd^4 \mf x \dd^4 \mf x'\; \Lambda(\mf x)\Lambda^*(\mf x')\mel{0}{\delta\hat{\rho}(\mf{x})\delta\hat{\rho}(\mf{x}')}{0}e^{-i\Omega(t-t')}\\
    &= \frac{\bar{g}_{ab}^2}{\hbar^2}  \left\langle\left(\int\dd^4 \mf x \Lambda(\mf x)e^{-i\Omega t}\delta\hat{\rho}(\mf{x})\right)\left(\int\dd^4 \mf x'\Lambda^*(\mf x')e^{i \Omega t'}{\delta\hat{\rho}(\mf{x}')}\right)\right\rangle\\
    &= \frac{\bar{g}_{ab}^2}{\hbar^2} \langle \delta \hat{\rho}(\Lambda^-)\delta\hat{\rho}(\Lambda^+)\rangle,
\end{align}
where $\Lambda(\mf x) = \chi(t) f(\bm x)$ is the spacetime smearing function of the interaction.

\section{Computing the $\mathcal{L}$ and $\mathcal{M}$ terms explicitly}\label{app:computations}

In the continuum limit (infinite condensate volume), the two-point function of the Bose-Einstein condensate can be written as

\begin{equation}~\label{eq:rhotwopt}
    \langle \delta\hat{\rho}(\mf x) \delta\hat{\rho}(\mf x')\rangle =  \frac{1}{g_{bb}} \frac{1}{(2\pi)^3} \int \dd^3 \bm k \frac{\hbar\omega_{\bm k}}{2} e^{\ii \mf k \cdot (\mf x - \mf x')},
\end{equation}
where $\bm k$ has units of inverse length, corresponding to the wavelengths of the density fluctuation field. We can then write the term $\mathcal{L}$ as

\begin{align}
    \mathcal{L}  &= \frac{\bar g_{ab}^2}{\hbar^2} \int \dd^4\mf x \dd^4 \mf x' \Lambda(\mf x) \Lambda(\mf x') e^{- i  \Omega(t-t')}\langle\delta\hat{\rho}(\mf x)\delta\hat{\rho}(\mf x')\rangle\\
    &= \frac{\bar g_{ab}^2}{g_{bb}\hbar }\frac{1}{(2\pi)^3}\int \dd^3 \bm k \frac{\omega_{\bm k}}{2} \int \dd^4\mf x \dd^4 \mf x'  e^{-i \omega_{\bm k}(t-t') + i \bm k \cdot (\bm x - \bm x')}e^{- i  \Omega(t-t')}\chi(t)\chi(t') f(\bm x) f(\bm x') \\
    &= \frac{\bar g_{ab}^2}{g_{bb}\hbar } \frac{1}{(2\pi)^3}\int \dd^3 \bm k \frac{\omega_{\bm k}}{2} |\tilde{\chi}(\omega_{\bm k}+\Omega)|^2 \tilde{f}(-\bm k) \tilde{f}(\bm k),
\end{align}
where
\begin{align}
    \tilde{\chi}(\omega) = \int \dd t \chi(t) e^{- i \omega t},\\
    \tilde{f}(\bm k) = \int \dd^3 \bm x f(\bm x) e^{- i \bm k \cdot \bm x},
\end{align}
and we used that the switching function $\chi(t)$ is real. For convenience we define the coupling constant $\bar{\lambda}$
\begin{equation}
    \bar{\lambda}^2 = \frac{\bar g_{ab}^2}{g_{bb}\hbar c_\tc{s}},
\end{equation}
use that $\omega_{\bm k} \approx c_\tc{s} |\bm k|$, and write the switching function as $\chi(t) = \beta(t/T)$, where $\beta$ determines the profile of the interaction and $T$ controls the interaction time. For instance, for a Gaussian pulse we have $\beta(t) = e^{- \frac{t^2}{2}}$ and $T$ corresponds to its standard deviation. We then obtain $\tilde{\chi}(\omega) = T \tilde{\beta}(\omega T)$ and
\begin{align}
    \mathcal{L} &= \bar{\lambda}^2 \frac{c_\tc{s}^2 T^2}{(2\pi)^3}\int \dd^3 \bm k \frac{|\bm k|}{2} |\tilde{\beta}(T(c_\tc{s}|\bm k|+\Omega))|^2 \tilde{f}(-\bm k) \tilde{f}(\bm k).
\end{align}
We now define the rescaled time and frequencies $\bar{T} = c_\tc{s} T$ and $\bar{\Omega} = \Omega/c_\tc{s}$ (with units of length and inverse length, respectively), allowing us to recast
\begin{equation}
    \mathcal{L} = \frac{\bar{\lambda}^2}{(2\pi)^3} \bar{T}^2 \int \dd^3 \bm k \frac{|\bm k|}{2} |\tilde{\beta}(\bar{T}(|\bm k|+\bar{\Omega}))|^2 \tilde{f}(-\bm k) \tilde{f}(\bm k).
\end{equation}
Importantly, the expression above with $\bar{\lambda}\mapsto\lambda/\sqrt{\hbar c}$, $\bar{T}\mapsto c T$, $\bar{\Omega} \mapsto \Omega/c$ corresponds to the excitation probability of a momentum-coupled UDW detector with switching function $\chi(t) = \beta(t/T)$. 

Specializing to the case where $\beta(t) = e^{-\frac{t^2}{2}}$ and the smearing functions are
\begin{equation}\label{eq:gaussSmearing}
    f(\bm x) = \frac{1}{(2\pi \sigma^2)^\frac{3}{2}}e^{- \frac{|\bm x|^2}{2\sigma^2}},
\end{equation}
we can integrate $\mathcal{L}$ in closed form:
\begin{align}
    \mathcal{L} &= \frac{\bar{\lambda}^2}{(2\pi)^2} \bar{T}^2 \int \dd^3 \bm k \frac{|\bm k|}{2} |e^{- \bar{T}^2(|\bm k| + \bar{\Omega})^2} e^{-\sigma^2 |\bm k|^2}= \frac{\bar{\lambda}^2}{2\pi} \bar{T}^2 \int \dd |\bm k| \,|\bm k|^3 e^{- \bar{T}^2(|\bm k| + \bar{\Omega})^2} e^{-\sigma^2 |\bm k|^2}\\
    &= \frac{\bar{\lambda}^2\bar{T}^2 e^{-\bar{T}^2 \bar{\Omega} ^2} \left(2 \sqrt{\sigma ^2+\bar{T}^2} \left(\sigma ^2+\bar{T}^4
   \bar{\Omega} ^2+\bar{T}^2\right)-\sqrt{\pi } \bar{T}^2 \bar{\Omega}  e^{\frac{\bar{T}^4 \bar{\Omega} ^2}{\sigma
   ^2+\bar{T}^2}} \left(2 \bar{T}^4 \bar{\Omega} ^2+3 \left(\sigma ^2+\bar{T}^2\right)\right)
   \text{erfc}\left(\frac{\bar{T}^2 \bar{\Omega} }{\sqrt{\sigma ^2+\bar{T}^2}}\right)\right)}{8
   \pi  \left(\sigma ^2+\bar{T}^2\right)^{7/2}}
\end{align}

An analogous reasoning can be made for the $\mathcal{M}$ term, although the computations are slightly more involved. Explicitly, we have
\begin{align}
    \mathcal{M}  &= -\frac{\bar g_{ab}^2}{\hbar^2} \int \dd V \dd V' \Lambda_\tc{a}(\mf x) \Lambda_\tc{b}(\mf x') e^{\ii \Omega(t+t')}\langle\mathcal{T}(\delta\hat{\rho}(\mf x)\delta\hat{\rho}(\mf x'))\rangle,\\
    &= -\frac{\bar g_{ab}^2}{\hbar^2} \int \dd V \dd V' \Lambda_\tc{a}(\mf x) \Lambda_\tc{b}(\mf x') e^{\ii \Omega(t+t')}(\langle\delta\hat{\rho}(\mf x)\delta\hat{\rho}(\mf x')\rangle\theta(t-t') +\langle\delta\hat{\rho}(\mf x')\delta\hat{\rho}(\mf x)\rangle\theta(t'-t)),\\
    &= -\frac{\bar g_{ab}^2}{g_{bb}\hbar } \frac{1}{(2\pi)^3} \int \dd^3 \bm k \frac{\omega_{\bm k}}{2} \int \dd^4\mf x \dd^4 \mf x'  \big(e^{-i \omega_{\bm k}(t-t') + i \bm k \cdot (\bm x - \bm x')}\theta(t-t')e^{ i  \Omega(t+t')}\chi(t)\chi(t') f(\bm x) f(\bm x' - \bm L)\\
    &\quad\quad\quad\quad\quad\quad\quad\quad\quad\quad\quad\quad\quad\quad\quad\quad\quad+e^{i \omega_{\bm k}(t-t') - i \bm k \cdot (\bm x - \bm x')}\theta(t'-t))e^{i  \Omega(t+t')}\chi(t)\chi(t') f(\bm x) f(\bm x' - \bm L)\big)\nonumber\\
     &= -\frac{\bar g_{ab}^2}{g_{bb}\hbar }\frac{1}{(2\pi)^3} \int \dd^3 \bm k \frac{\omega_{\bm k}}{2} \int \dd t \dd t'  \big(e^{-i \omega_{\bm k}(t-t')} \theta(t-t') e^{-i \bm k \cdot \bm L}+e^{i \omega_{\bm k}(t-t') }\theta(t'-t) e^{i \bm k \cdot \bm L})\chi(t)\chi(t')e^{i  \Omega(t+t')} \tilde{f}(-\bm k)\tilde{f}(\bm k)\nonumber\\
     &= -\frac{\bar g_{ab}^2}{g_{bb}\hbar } \frac{1}{(2\pi)^3}\int \dd^3 \bm k\, \frac{\omega_{\bm k}}{2}  2\int \dd t \dd t'  \chi(t)\chi(t') e^{i  \Omega(t+t')}e^{-i \omega_{\bm k}(t-t')}\theta(t-t')   e^{-i \bm k \cdot \bm L} \tilde{f}(-\bm k)\tilde{f}(\bm k)\nonumber,
\end{align}
where in the last equality we made the changes of variables $t\leftrightarrow t'$ and $\bm k \mapsto -\bm k$ in the second summand. We now define
\begin{equation}
    Q_\chi(\omega_{\bm k}, \Omega, T) = \int \dd t \dd t'  \chi(t)\chi(t') e^{i  \Omega(t+t')}e^{-i \omega_{\bm k}(t-t')}\theta(t-t'),
\end{equation}
and again assume that $\chi(t) = \beta(t/T)$ and $\omega_{\bm k} = c_\tc{s}|\bm k|$, so that
\begin{align}
    Q_\chi(c |\bm k|, \Omega, T) &= \int \dd t \dd t'  \beta(t/T)\beta(t'/T) e^{i  \Omega(t+t')}e^{-i c_\tc{s} |\bm k| (t-t')}\theta(t-t')\\
    &= T^2 \int \dd u \dd u'  \beta(u)\beta(u') e^{i  \Omega T(u+u')}e^{-i c_\tc{s} T |\bm k| (u-u')}\theta(u-u')\\
    & = T^2 Q_\beta(|\bm k|, \Omega/c_\tc{s}, c_\tc{s} T).
\end{align}
Plugging this result into the expression for $\mathcal{M}$ and using again $\bar{T} = c_\tc{s} T$ and $\bar{\Omega} = \Omega/c_\tc{s}$ yields
\begin{align}
    \mathcal{M} &= -\bar{\lambda}^2  \frac{\bar{T}^2}{(2\pi)^3}\int \dd^3 \bm k \frac{|\bm k|}{2} 2Q_\beta(|\bm k|, \bar{\Omega},\bar{T})   e^{-i \bm k \cdot \bm L} \tilde{f}(-\bm k)\tilde{f}(\bm k).
\end{align}
Once again, the expression above with $\bar{\lambda}\mapsto\lambda/\sqrt{\hbar c}$, $\bar{T}\mapsto c T$, $\bar{\Omega} \mapsto \Omega/c$ corresponds to the $\mathcal{M}$ term of a momentum-coupled UDW detector with switching function $\chi(t) = \beta(t/T)$.

Specializing to the case where $\beta(t) = e^{-\frac{t^2}{2}}$ we find a closed-form expression for $Q(|\bm k|,\bar{\Omega},\bar{T})$~\cite{Pozas-Kerstjens:2015,boris}:
\begin{equation}
    2Q_\beta(|\bm k|,\bar{\Omega},\bar{T}) = 2\pi e^{-(\bar{\Omega}^2+|\bm k|^2)\bar{T}^2}(1 - i  \text{erfi}(|\bm k| \bar{T})).
\end{equation}
Using the smearing functions as in Eq.~\eqref{eq:gaussSmearing} we can write
\begin{align}
    \mathcal{M} &= -\bar{\lambda}^2 \frac{ \bar{T}^2}{(2\pi)^2}\int \dd^3 \bm k \frac{|\bm k|}{2} e^{-(\bar{\Omega}^2+|\bm k|^2)\bar{T}^2}(1 - i  \text{erfi}(|\bm k| \bar{T}))   e^{-i \bm k \cdot \bm L} e^{- \sigma^2 |\bm k|^2}\\
    &= -\bar{\lambda}^2 \frac{ \bar{T}^2}{2\pi}\int \dd |\bm k| \dd\theta \sin\theta \frac{|\bm k|^3}{2} e^{-(\bar{\Omega}^2+|\bm k|^2)\bar{T}^2}(1 - i  \text{erfi}(|\bm k| \bar{T}))   e^{-i |\bm k| |\bm L| \cos\theta} e^{- \sigma^2 |\bm k|^2}\\
    &= -\bar{\lambda}^2 \frac{ \bar{T}^2}{(2\pi)}\int \dd |\bm k| |\bm k|^3 e^{-(\bar{\Omega}^2+|\bm k|^2)\bar{T}^2}(1 - i  \text{erfi}(|\bm k| \bar{T})) \text{sinc}(|\bm k| L) e^{- \sigma^2 |\bm k|^2}\\
    &= -\bar{\lambda}^2 \frac{ \bar{T}^2}{2\pi |\bm L|}\int \dd |\bm k| |\bm k|^2 e^{-(\bar{\Omega}^2+|\bm k|^2)\bar{T}^2}(1 - i  \text{erfi}(|\bm k| \bar{T})) \text{sin}(|\bm k| L) e^{- \sigma^2 |\bm k|^2}.
\end{align}
Although at first unhinged, the integral above can be computed in closed-form. First, define
\begin{align}
    \mathcal{M}^+ &= -\bar{\lambda}^2 \frac{ \bar{T}^2}{2\pi |\bm L|}\int \dd |\bm k| |\bm k|^2 e^{-(\bar{\Omega}^2+|\bm k|^2)\bar{T}^2}\text{sin}(|\bm k| L) e^{- \sigma^2 |\bm k|^2}\\
    \mathcal{M}^- &= i  \bar{\lambda}^2 \frac{ \bar{T}^2}{2\pi |\bm L|}\int \dd |\bm k| |\bm k|^2 e^{-(\bar{\Omega}^2+|\bm k|^2)\bar{T}^2}\text{sin}(|\bm k| L) e^{- \sigma^2 |\bm k|^2}\text{erfi}(|\bm k| \bar{T}),
\end{align}
so that $\mathcal{M} = \mathcal{M}^+ + \mathcal{M}^-$. The first integral involves only complex exponentials, Gaussians and polynomials and can be solved using standard methods:
\begin{equation}
    \mathcal{M}^+ = \frac{\bar{\lambda}^2\bar{T}^2 e^{-\bar{T}^2 \bar{\Omega} ^2} \left( \sqrt{\pi } e^{-\frac{L^2}{4
   \left(\sigma ^2+\bar{T}^2\right)}} \left(L^2-2 \left(\sigma ^2+\bar{T}^2\right)\right)
   \text{erfi}\left(\frac{L}{2 \sqrt{\sigma ^2+\bar{T}^2}}\right)-2L \sqrt{\sigma
   ^2+\bar{T}^2}\right)}{16 \pi  L \left(\sigma ^2+\bar{T}^2\right)^{5/2}}.
\end{equation}
The integral defining $\mathcal{M}^-$ is less trivial. To solve it we first state the result shown in~\cite{closedform2024}:
\begin{equation}
    \int \dd |\bm k| e^{- (\bar{T}^2 + \sigma^2)|\bm k|^2}\sin(|\bm k| L) \text{erfi}(|\bm k| \bar{T}) = \frac{e^{- \frac{L^2}{4(\bar{T}^2 + \sigma^2)}}}{\sqrt{\bar{T}^2 + \sigma^2}}\frac{\sqrt{\pi}}{2}\text{erf}\left(\frac{L \bar{T}}{2 \sqrt{\sigma^2}\sqrt{\bar{T}^2 + \sigma^2}}\right).
\end{equation}
We can now recast the integral of $\mathcal{M}^-$ as
\begin{align}
    \mathcal{M}^- &= i  \bar{\lambda}^2 \frac{ \bar{T}^2e^{- \bar{\Omega}^2 \bar{T}^2}}{2\pi |\bm L|} \int \dd |\bm k| |\bm k|^2  e^{- (\bar{T}^2 + \sigma^2)|\bm k|^2}\sin(|\bm k| L) \text{erfi}(|\bm k| \bar{T})\\
    & = i  \bar{\lambda}^2 \frac{ \bar{T}^2e^{- \bar{\Omega}^2 \bar{T}^2}}{2\pi |\bm L|} \int \dd |\bm k| \left(- \dv{}{\sigma^2}  e^{- (\bar{T}^2 + \sigma^2)|\bm k|^2}\right)\sin(|\bm k| L) \text{erfi}(|\bm k| \bar{T})\\
    & = i  \bar{\lambda}^2 \frac{ \bar{T}^2e^{- \bar{\Omega}^2 \bar{T}^2}}{2\pi |\bm L|}\left( - \dv{}{\sigma^2}\left(\int \dd |\bm k|   e^{- (\bar{T}^2 + \sigma^2)|\bm k|^2}\sin(|\bm k| L) \text{erfi}(|\bm k| \bar{T})\right)\right)\\
    & = i  \bar{\lambda}^2 \frac{ \bar{T}^2e^{- \bar{\Omega}^2 \bar{T}^2}}{2\pi |\bm L|} \left(- \dv{}{\sigma^2}\left(\frac{e^{- \frac{L^2}{4(\bar{T}^2 + \sigma^2)}}}{\sqrt{\bar{T}^2 + \sigma^2}}\frac{\sqrt{\pi}}{2}\text{erf}\left(\frac{L \bar{T}}{2 \sqrt{\sigma^2}\sqrt{\bar{T}^2 + \sigma^2}}\right)\right)\right).
\end{align}
The operation above then gives us
\begin{equation}
    \mathcal{M}^- = -\frac{i \bar{\lambda}^2\bar{T}^2 e^{-\bar{T}^2 \bar{\Omega} ^2}\left(\sqrt{\pi }  e^{-\frac{L^2}{4 \left(\sigma ^2+\bar{T}^2\right)}} \left(L^2-2 \left(\sigma ^2+\bar{T}^2\right)\right)
   \text{erf}\left(\frac{L \bar{T}}{2 \sigma  \sqrt{\sigma ^2+\bar{T}^2}}\right)-2 L \bar{T}
  e^{-\frac{L^2}{4
   \sigma ^2}} \sqrt{1+\tfrac{\bar{T}^2}{\sigma^2}} \left(2+\tfrac{\bar{T}^2}{\sigma^2}\right)\right) }{16 \pi  L  \left(\sigma
   ^2+\bar{T}^2\right)^{5/2}}.
\end{equation}

\section{Signalling between the probes and relativistic behaviour}\label{app:checks}

We can quantify how much of the entanglement between the probes is due to signaling by using the signaling estimator defined in~\cite{ericksonNew}:
\begin{equation}
        \mathcal{I} = \begin{cases}
            \displaystyle{\frac{\mathcal{N}^-}{\mathcal{N}}}& \mathcal{N}>0\\[2mm]
            0  & \mathcal{N} = 0,
        \end{cases}
\end{equation}
where $\mathcal{N}^- = \text{max}(0,|\mathcal{M}^-| - \mathcal{L})$, where $\mathcal{M}^-$ corresponds to the causal exchanges encoded in the $\mathcal{M}$ term, and is defined in Section~\ref{app:computations}. This estimator satisfies $0\leq \mathcal{I}\leq 1$ such that $\mathcal{I} = 0$ implies that the entanglement comes entirely from quantum fluctuations of the field (encoded in the term $\mathcal{M}^+$), and $\mathcal{I} = 1$ corresponds to the case where all entanglement acquired by the probes is due to communication through the field. 

In Figure~\ref{fig:signaling} we plot the signaling estimator as a function of the frequency of the trapping potential $\Omega$ for the setups considered in Figure 3 of the main text, when two potassium probes couple to a rubidium condensate for times of the order of $\text{ms}$ while separated by distances $L = 4.6 c_\tc{s} T$, of the order of $1.25 \mu \text{m}$. In the setups considered, the signaling estimator stays at exactly zero for sufficiently small values of the trap frequency $\Omega$, and sharply raise to their maxima, which remains below 75\% for all $\Omega$. For the values of $\Omega$ shown as solid lines in Fig. 3 of the main text we have $\mathcal{I}$ identically zero, implying that these cases correspond to genuine entanglement harvesting, where communication between the probes does not contribute to their entanglement.

\begin{figure}[h!]
    \centering
    \includegraphics[width=0.85\linewidth]{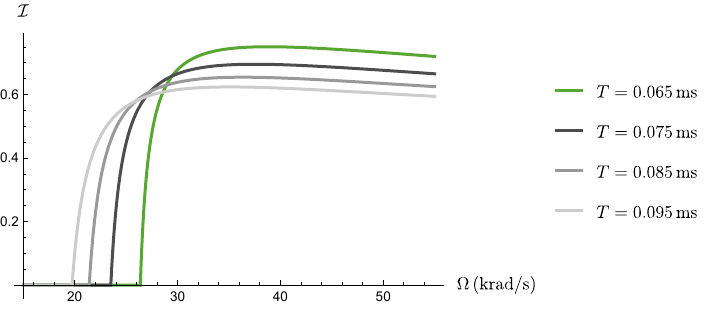}
    \caption{Plot of the signaling estimator as a function of $\Omega$ for the different setups considered in Fig. 2 of the main text when potassium probes coupled to a rubidium BEC are separated by distances $L = 5.25 c_\tc{s} T$. }
    \label{fig:signaling}
\end{figure}

The density fluctuations behave relativistically for wavenumbers with $|\bm k|\leq \xi^{-1}$, where $\xi$ is the healing length. We can estimate the wavenumbers relevant in the protocol of entanglement harvesting by writing the relevant terms $\mathcal{L}$, $\mathcal{M}^+$ and $\mathcal{M}^-$ as an integral in $|\bm k|$:
\begin{align}
    \mathcal{L} &= \int \dd |\bm k|\, \mathcal{L}(|\bm k|)\\
    \mathcal{M}^+ &= \int \dd |\bm k|\, \mathcal{M}^+(|\bm k|)\\
    \mathcal{M}^- &= \int \dd |\bm k| \,\mathcal{M}^-(|\bm k|).
\end{align}
We plot these functions as a function of $|\bm k|$ for the protocol utilizing potassium impurities coupled to a rubidium BEC ($\xi \approx 123\, \text{nm}$) in Fig.~\ref{fig:healing} when the interaction times are controlled by the parameter $T = 0.065 \,\text{ms}$, the probes are separated by $L = 4.6 c_\tc{s} T$ and the frequency of the harmonic traps correspond to the negativity peak $\Omega = 26.3\,\text{krad/s}$. We see that the relevant terms for genuine entanglement harvesting, $\mathcal{L}(|\bm k|)$ and $\mathcal{M}^+(|\bm k|)$ are well within the relativistic regime, with the integrands only being relevant for frequencies smaller than the healing length. Although the tails of $\mathcal{M}^-(|\bm k|)$ extend beyond $\xi^{-1} \approx 8.13\, (\text{nm})^{-1}$, less than 4\% of its total integral is contained in the region $|\bm k|>\xi^{-1}$.

\begin{figure}[h!]
    \centering
    \includegraphics[width=0.85\linewidth]{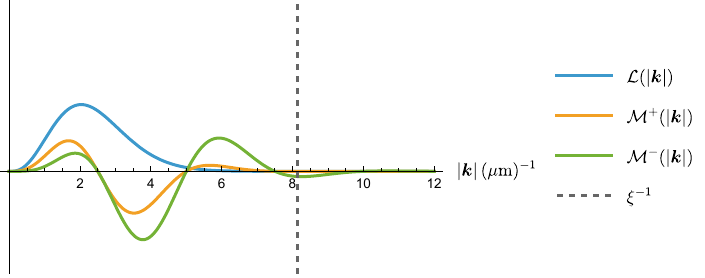}
    \caption{Plot of the integrands $\mathcal{L}(|\bm k|)$, $\mathcal{M}^+(|\bm k|)$, and $\mathcal{M}^-(|\bm k|)$ as a function of $|\bm k|$ for two potassium impurities probing a rubidium BEC while separated by a distance $L = 4.6 c_\tc{s} T$, where $T = 0.065 \,\text{ms}$ controls the time profile of the interaction, and the trap frequency of the impurities is $\Omega = 26.3\, \text{krad}/\text{s}$}.
    \label{fig:healing}
\end{figure}

\section{Measuring the $\mathcal{M}$ term}\label{app:Mterm}

In this section we will discuss the local operations that have to be applied to the two-impurity density operator after the short time interaction with the field that allows for computation of the negativity of its final state. Given the density operator
\begin{equation}
    \hat{\rho}_{\tc{ab}} = \begin{pmatrix}
        1 - \mathcal{P}_\tc{aa} - \mathcal{P}_\tc{bb} & 0 & 0 & \mathcal{M}^*\\
        0 & \mathcal{P}_{\tc{bb}} & \mathcal{P}_{\tc{ab}} & 0\\
        0 & \mathcal{P}_{\tc{ab}}^* & \mathcal{P}_\tc{aa} & 0\\
        \mathcal{M} & 0 & 0 & 0
    \end{pmatrix} + \mathcal{O}(g_{ab}^4),
\end{equation}
one can access the $\mathcal{L}$ term by measuring whether the impurity became excited after the interactions, with $P_e = \mathcal{L}$ for each of the identical qubits. However, the $\mathcal{M}$ term requires additional steps. Explicitly, we can recast it in terms of expectation values of correlations of different spin operators:
\begin{align}
    \Re(\mathcal{M}) = \frac{1}{4}(\langle\hat{\sigma}_x\otimes\hat{\sigma}_x \rangle - \langle\hat{\sigma}_y\otimes\hat{\sigma}_y \rangle),\\
    \Im(\mathcal{M}) = -\frac{1}{4}(\langle\hat{\sigma}_x\otimes\hat{\sigma}_y \rangle + \langle\hat{\sigma}_y\otimes\hat{\sigma}_x \rangle).
\end{align}
The quantities above can then be obtained in terms of mutual excitation probabilities of both probes after performing specific unitaries prior to detection. Explicitly, denote by $P_{ij}(\hat{U}_\tc{a},\hat{U}_\tc{b}) = \text{Tr}(\hat{U}_\tc{a}\otimes\hat{U}_\tc{b}\, \hat{\rho}_\tc{ab}\, \hat{U}_\tc{a}^\dagger \otimes\hat{U}_\tc{b}^\dagger \ket{i}\!\!\bra{i}\otimes\ket{j}\!\!\bra{j})$ for $i,j\in\{g,e\}$ and local unitaries $\hat{U}_\tc{a}$, $\hat{U}_\tc{b}$. $P_{ij}(\hat{U}_\tc{a},\hat{U}_\tc{b})$ then denotes the probability that the impurities are measured in the state $\ket{ij}$ after the local unitaries $\hat{U}_\tc{a}$ and $\hat{U}_\tc{b}$ are applied to the systems. Let $\hat{H}$ denote the Hadamard gate with $\hat{H}\ket{g} = \frac{1}{\sqrt{2}}(\ket{g}+\ket{e})$, $\hat{H}\ket{e} = \frac{1}{\sqrt{2}}(\ket{g}-\ket{e})$ and $\hat{S}$ denote the phase shift $\hat{S}\ket{g} = \ket{g}$ and $\hat{S}\ket{e} = -\ii \ket{e}$. Then we can write
\begin{align}
    \langle\hat{\sigma}_x\otimes\hat{\sigma}_x \rangle &= P_{gg}(\hat{H},\hat{H})+P_{ee}(\hat{H},\hat{H}) - P_{ge}(\hat{H},\hat{H})-P_{eg}(\hat{H},\hat{H}),\\
    \langle\hat{\sigma}_y\otimes\hat{\sigma}_y \rangle &= P_{gg}(\hat{H}\hat{S},\hat{H}\hat{S})+P_{ee}(\hat{H}\hat{S},\hat{H}\hat{S}) - P_{ge}(\hat{H}\hat{S},\hat{H}\hat{S})-P_{eg}(\hat{H}\hat{S},\hat{H}\hat{S}),\\
    \langle\hat{\sigma}_x\otimes\hat{\sigma}_y \rangle &= P_{gg}(\hat{H},\hat{H}\hat{S})+P_{ee}(\hat{H},\hat{H}\hat{S}) - P_{ge}(\hat{H},\hat{H}\hat{S})-P_{eg}(\hat{H},\hat{H}\hat{S}),\\
    \langle\hat{\sigma}_y\otimes\hat{\sigma}_x \rangle &= P_{gg}(\hat{H}\hat{S},\hat{H})+P_{ee}(\hat{H}\hat{S},\hat{H}) - P_{ge}(\hat{H}\hat{S},\hat{H})-P_{eg}(\hat{H}\hat{S},\hat{H}).
\end{align}


For the experimental implementation of $\hat{H}$ and $\hat{H}\hat{S}$ the traps are modulated in position (shaking) allowing the realization of the corresponding gates as demonstrated in e.g. \cite{MO_Scelle2013MotionalCoherence}. The additional phase shift $\hat{S}$ is achieved by controlling the relative phase shift of the modulations of the two traps---a $\pi/2$ phase difference in the modulation implements the necessary phase shift in $\hat{H}\hat{S}$.

\section{Infidelity estimates for concrete setup }\label{app:experimental}

There are many different ways for generating traps for neutral atoms. For the discussion of systematic infidelities of the unitary transformations e.g. Hadamard gates for motional degrees of freedom, or readout fidelities, we have chosen a concrete setup based on optical lattices (see Fig.~\ref{fig:potentials}). The corresponding interfering light beams are realized at the tuneout wavelength of $790 \text{nm}$ implementing a potential for the impurities but leaving the density of the background BEC unaffected. We would like to emphasize that the following study should be understood as a proof-of-concept rather than a detailed study of experimental possibilities. There is plenty of room for improvement employing state-of-the-art optimal quantum control algorithms. 

For the realization of the harmonic traps with adjustable trap frequency $\Omega$, we employ an optical lattice i.e.\ interfering laser beams leading to a periodic light shift potential $V(x)=V_0 \cos(x)$ \cite{MO_lattices_RevModPhys}. Since the laser frequency is fixed by the tuneout wavelength, the periodic potential with a periodicity of $1 \mu \text{m}$ has to be realized by two laser beams which enclose an angle of $46^\circ$. The resulting potential (Unruh-DeWitt detector trap potential) and the local states are shown in the upper row in Fig.~\ref{fig:potentials}. 
\begin{figure}[h!]
    \centering
    \includegraphics[width=0.85\linewidth]{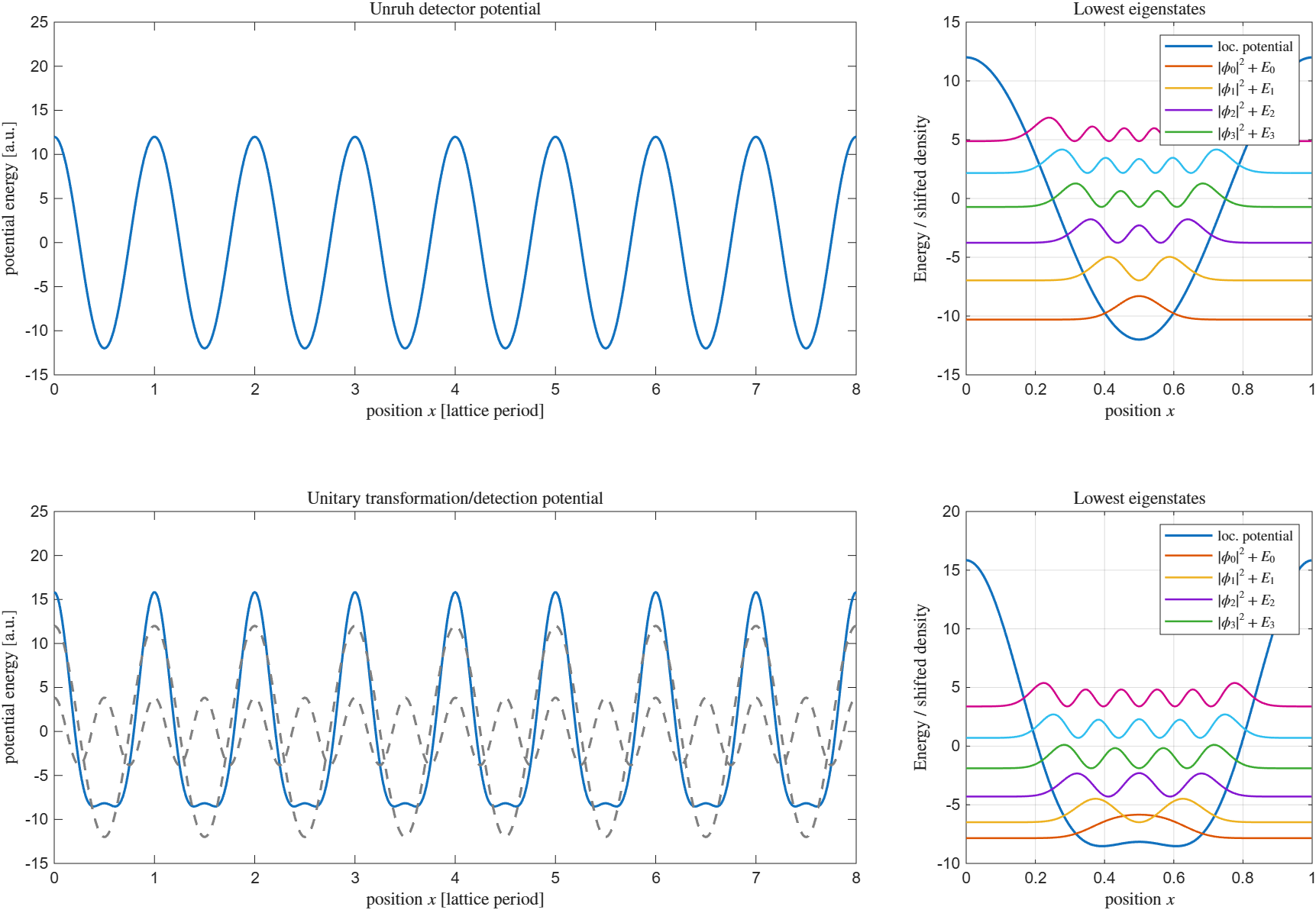}
    \caption{The potentials used for implementation of entanglement harvesting of a BEC. Upper row: A periodic potential (UDW detector trapping potential) realized at the tuneout wavelength for Rb generates traps for the impurities with a distance of $1 \mu \text{m}$. The corresponding lowest eigenstates localized at a potential minimum are in good approximation harmonic oscillator states (localized UDW detectors). Lower row: For unitary transformations as well as for detection of populations with high fidelity (low infidelity) the local traps are modified with an additional periodic potential with twice the spatial frequency. The corresponding eigenstates and eigenenergies are depicted.}.
    \label{fig:potentials}
\end{figure}

The implementation of high fidelity unitaries for the two lowest energy states of the UDW detector potential can be achieved by making the potential anharmonic. We choose to add an additional periodic potential to generate local double-well structure---for this discussion, $V(x)=V_0 (\cos(x) + \frac{1}{3} \cos(2x))$, see Fig.~\ref{fig:potentials}. This makes the energy difference of the two lowest states smaller than the other splittings and additionally allows the straightforward implementation of the necessary unitaries by periodically modulating the position of the additional potential. In the following we will refer to this additional potential as the manipulation/detection potential.

For the infidelity study we have implemented numerically the single particle dynamics in potentials of the general form $V(x)=V_0 (\cos(x)+\alpha(t) \cos(2 x))$.  The evolution of the populations of the corresponding eigenstates as the potential is changed $\alpha(t) = \frac{1
}{3}(t/T)^{7/6}$ is shown in Fig.~\ref{fig:adiabticloadingofdetectionpotential}. The timescale is given in units of the oscillation timescale of the local harmonic trap at $t=0$. Clearly the infidelity is well below $10^{-5}$. The chosen exponent of the ramp mainly reduced the $t=0$ population in the second excited state below $10^{-7}$. Employing optimal control this infidelity can be further reduced.

\begin{figure}[h!]
    \centering
    \includegraphics[width=0.75\linewidth]{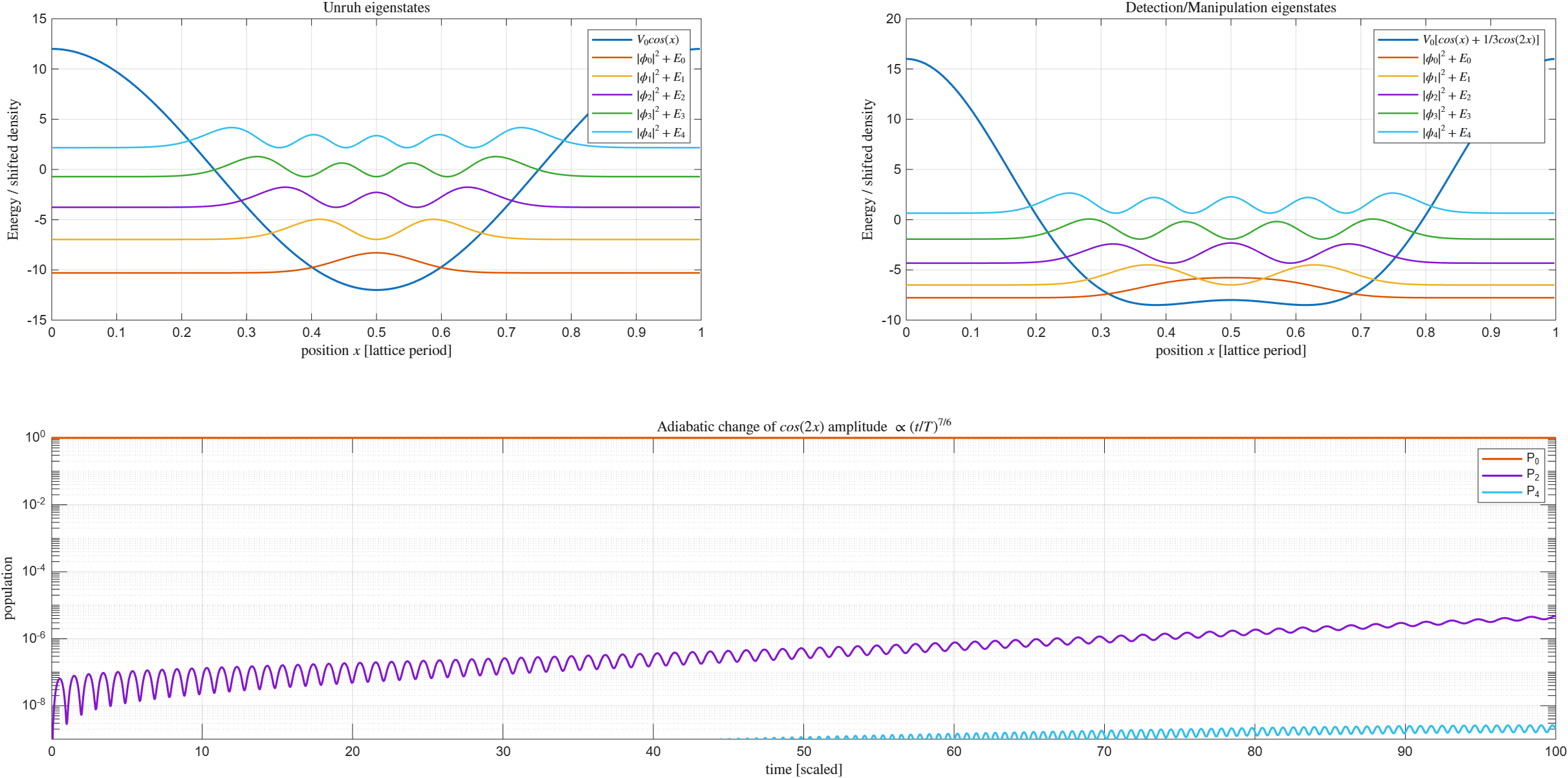}
    \caption{The change from UDW detector potential to manipulation/detection potential. The upper panels represent the corresponding potentials, eigenstates and eigenenergies for the initial and final potential. The lower panel shows the evolution of the populations of the eigenstates at the corresponding potential within the ramp $(t/T)^{7/6}$. Due to the symmetry of the situation only the second excited states and higher even states can be excited.}
\label{fig:adiabticloadingofdetectionpotential}
\end{figure}

Given that the impurities are transferred into the double-well geometry, the further manipulations rely on the additional control of the relative position of the two periodic potentials. This can be achieved by changing the potential periodically in time as $V(x, t) = V_0 (\cos(x)+\frac{1}{3} \cos(2 (x-x_t))$, for $x_t = A\sin \small{(}\tilde{\Omega}t\small{)}$ with $A=0.0016$ and the frequency $\tilde{\Omega}$ corresponding to the energy difference between ground and first excited state in the manipulation/detection potential, which is different to the Unruh detector potential characterized by $\Omega$. The results are shown in Fig.~\ref{fig:implementationofHadmar} and reveal that an infidelity well below $10^{-5}$ is possible with this realization, thus allowing for the necessary Hadamard transformations at the interaction time leading to a 50-50 superposition.  
\begin{figure}[h!]
    \centering
    \includegraphics[width=0.75\linewidth]{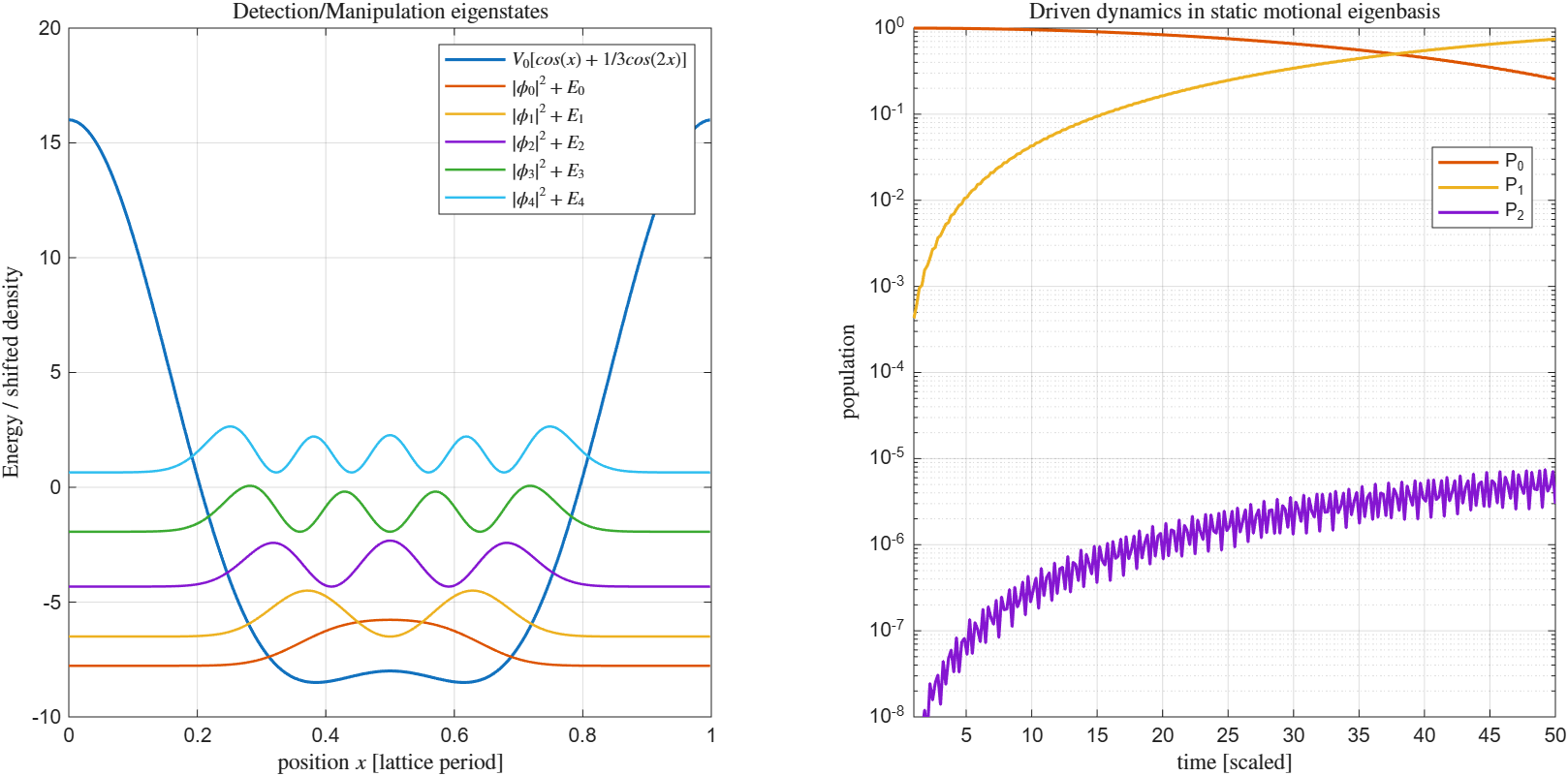}
    \caption{Periodically modulating the relative position between the two lattice potentials allows the controlled coupling of ground and first excited state. Left panel: the eigenstates and eigenenergies for the Hadamard transformation. Right panel: the evolution of the populations of the eigenenergies as a function of time for given driving (with parameters as in the text). The infidelity is given by the residual excitation of the second excited state which stays well below $10^{-5}$ level for the necessary rotation leading to a 50-50 superposition. The time is scaled to the energy difference between ground and first excited state.}
\label{fig:implementationofHadmar}
\end{figure}

The final step is the readout of the population of the ground and the first excited state --- either without unitary transformation (to estimate $\mathcal{P}$) or after the discussed unitary transformation (to extract the information about $\mathcal{M}$). For the population readout the double well geometry is again employed which allows the control of the barrier height as well as the symmetry (tilt) of the potential. The chosen parameter change maps the ground state to the right well of the tilted double potential, while the population of the left well is directly connected to the initial excited state population. Here again, we rely on adiabaticity during the change of the potential parameter. Non-adiabatic transitions limit the achievable fidelity. The results for the specifically chosen ramps  $x_t=0.075(t/T)$ and $\alpha(t)=1/3+1.2(t/T)^2$ are shown in Fig.~\ref{fig:implementationofpopulationreadout}. The left panel in the upper row shows the situation prior to the readout (discussed before), the right panel displays the final eigenstates in the tilted double well potential. The panel in the middle row shows the probability to find the particle in the corresponding eigenstate, which changes due to non-adiabatic transitions between the states during the parameter ramps. The lowest panel gives the probability to detect the particle after the ramp either on the left or right side. Clearly the infidelity is below $10^{-5}$ and thus is not limiting the estimation of populations $\mathcal{P}$ and combined with the prior discussed unitaries for the estimation of $\mathcal{M}$. Thus, negativity values in the order of $10^{-4}$ appear experimentally detectable.

These are estimations of well defined systematic errors. Only the experimental implementation of this scheme will reveal if this limit can be reached in a real experiment. We would like to emphasize that no optimization algorithms have been applied to reach this performance. A detailed study is beyond the scope of this paper and the given discussion should be seen as indicative that experimental observation of entanglement harvesting with BECs is within reach.

\begin{figure}[h]
    \centering
    \includegraphics[width=0.75\linewidth]{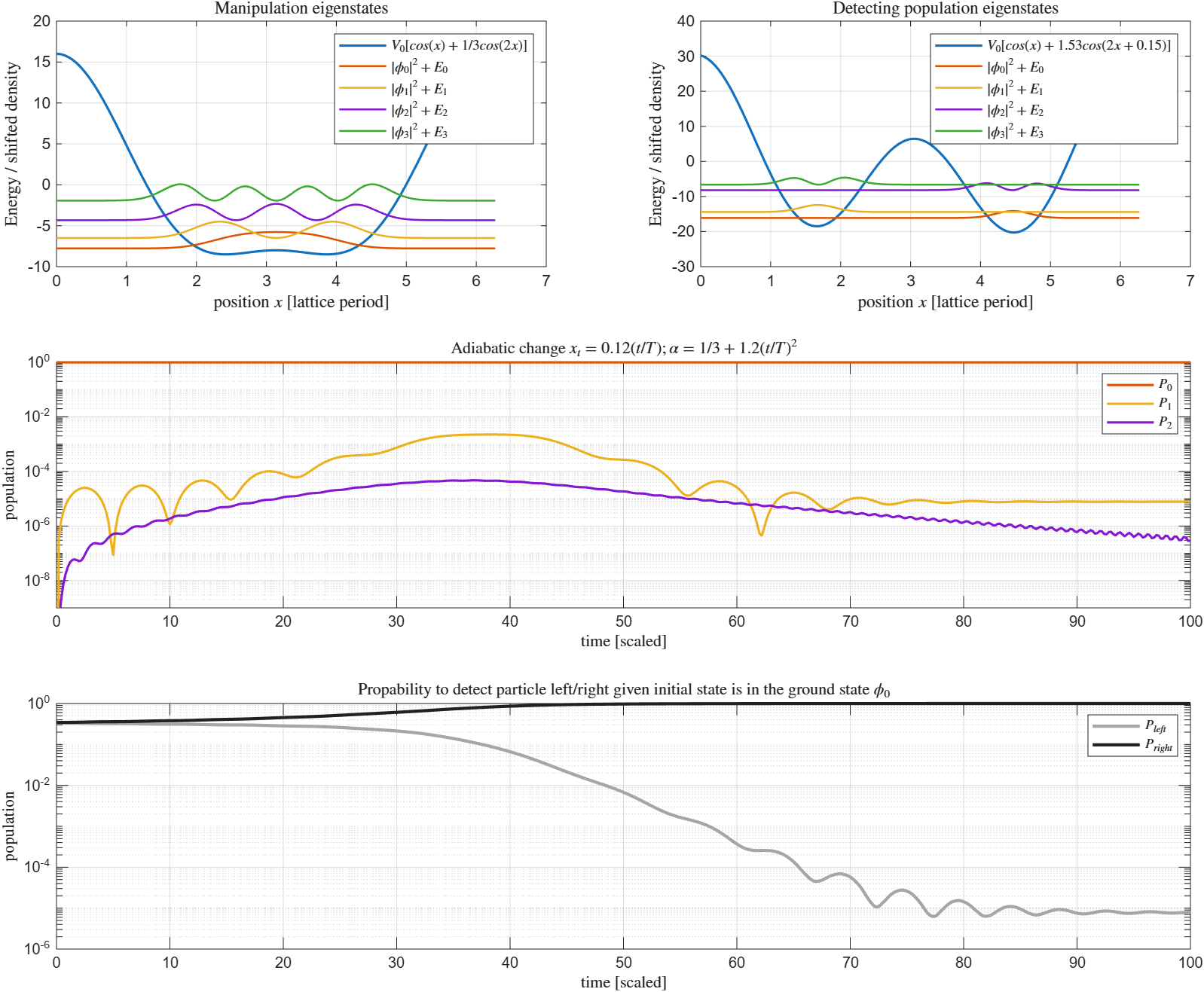}
    \caption{Detection of excited state amplitude via adiabatic mapping to a tiled double well potential. Upper panels: left - eigen states and eigen energies in the manipulation/detection potential; right - the final states after employing the ramp of potential height and tilt as detailed in the text. Middle panel:  the population of the instant eigenstates during the ramp. The population is changing due to non-adiabatic transitions. Lower panel: The integrated probability to detect a particle in left or right of the barrier. Clearly, this method allows the detection of the ground state amplitude with an infidelity below $10^{-5}$.}
\label{fig:implementationofpopulationreadout}
\end{figure}

\end{document}